\documentclass[a4paper,11pt]{article}
\pdfoutput=1 
\usepackage{graphicx}  
\usepackage{dcolumn}   
\usepackage{bm,relsize}        
\usepackage{amssymb, amsmath}
\usepackage{textcomp}
\usepackage{wasysym}
\usepackage{slashed}
\usepackage{caption, subcaption}
\usepackage{multirow}
\usepackage[para,online,flushleft]{threeparttable}

\usepackage{lipsum, color}
\usepackage[usenames,dvipsnames,svgnames]{xcolor}
\usepackage{braket}
\usepackage{jheppub} 
\usepackage{booktabs}
\usepackage{pdflscape}
\usepackage[utf8]{inputenc}

\def\beq{\begin{equation}}
\def\eeq{\end{equation}}
 \def\be{\begin{equation}} \def\ee{\end{equation}}
\def\bea{\begin{eqnarray}} \def\eea{\end{eqnarray}}

\newcommand{\mr}{m_r}
\def\mphi{m_\phi}

\usepackage{caption}
\usepackage{subcaption}
\usepackage{multirow}
\usepackage[normalem]{ulem}
\allowdisplaybreaks

\def\kslash{\rlap{\hspace{0.02cm}/}{k}}
\def\lslash{\rlap{\hspace{-0.02cm}/}{l}}

\newcommand{\ReducedG}{\tilde{G}}
\renewcommand{\Im}{\operatorname{Im}}

\def\coneprime{c_{\langle r^2\rangle}}
\def\ctwoprime{c_{E^2}}
\def\cthreeprime{c_{B^2}}
\def\cfourprime{c_{\langle r^4\rangle}}

\def\d6tilde{\tilde{d}_6}

\begin{document}

\title{Effective Field Theory  Perspective On  King Non-linearity}

\author[a]{Beno\^{i}t Assi,}
\author[b]{Sam Carey,}
\author[c]{Sebastian J\"ager,}
\author[d]{Gabriel Lee,}
\author[b]{Gil Paz,}
\author[e]{Gilad Perez,}
\author[a]{and Jure Zupan}

\affiliation[a]{Department of Physics, University of Cincinnati, Cincinnati, Ohio 45221, USA}
\affiliation[b]{Department of Physics and Astronomy, Wayne State University, Detroit, Michigan 48201, USA}
\affiliation[c]{University of Sussex, Department of Physics and Astronomy, Falmer, Brighton BN1 9QH, UK}
\affiliation[d]{Lyft, Inc., Toronto, Ontario M5H 4A6 Canada}
\affiliation[e]{Department of Particle Physics and Astrophysics, Weizmann Institute of Science, Rehovot 7610001, Israel}

\emailAdd{assibt@ucmail.uc.edu}
\emailAdd{samcarey@wayne.edu}
\emailAdd{S.Jaeger@sussex.ac.uk}
\emailAdd{gabr.lee@gmail.com}
\emailAdd{gilad.perez@weizmann.ac.il}
\emailAdd{gilpaz@wayne.edu}
\emailAdd{zupanje@ucmail.uc.edu}

\date{\today}

\preprint{WSU-HEP-2503}

\abstract{
Precision spectroscopic measurements of isotope shifts have recently reached  a high level of accuracy. Tests of King non-linearity (NL) along isotope chains have been proposed as a tool to search for fifth-force mediators. At the same time, these tests can potentially teach us about the structure of heavy nuclei at unprecedented precision, where King NL has already been observed in several systems.
A robust interpretation of the existing data, however, is hampered by incomplete control over the Standard Model (SM) contributions. We develop a systematic effective field theory framework, matching the SM onto scalar non-relativistic QED in the infinite nuclear mass limit and then onto quantum-mechanical potentials. This approach organizes all nuclear effects into a small set of Wilson coefficients and cleanly separates short- and long-distance physics. We show that the commonly used treatment of the $\langle r^2\rangle^2$ term needs to be reconsidered, as it arises only at second-order in perturbation theory, and we derive the long-range $1/r^4$ potential from nuclear polarizability. Applying the framework to hydrogen-like systems, we provide a transparent classification of SM sources of King NL relevant for current and future isotope-shift experiments. The formalism can be applied to learn about the shape of the heavy scalar nuclei at a higher level of precision and detail 
than
what was previously attainable. 
}

\maketitle

\section{Introduction}
 Precision spectroscopy of isotope shifts has long been a powerful tool for probing nuclear structure, such as determining the nuclear charge radii, and searching for physics beyond the Standard Model (SM)~\cite{Mack:1956xe,King1984, Fortson1990_NuclearStructurePNC}. 
More recently, 
 {\em ``King non-linearity''} (King NL) has been proposed as a novel method to constrain the existence of new light mediators that would induce a fifth force between the nucleus and electrons~\cite{Delaunay:2016brc,Berengut:2017zuo}.  To measure King NL, one compares frequency differences, $\Delta \nu_i^{a}$, between two or more atomic transitions ($i$) across a chain of at least four isotopes ($a$). 
If only the leading mass-shift and field-shift effects are present, these frequency differences are expected to lie on a straight line when plotted against mass shifts, $\mu_a=1/m_{A_a}-1/m_{A_a'}$, where $m_{A_a},m_{A_a'}$ are the nuclear masses of the isotopes \cite{King:1963}.  
The key insight proposed in refs.~\cite{Berengut:2017zuo,Delaunay:2016brc} was that since the fifth-force mediators induce calculable King NL, the absence of such effects can be used to set bounds on the couplings and mass of the mediator.
As long as the observed $\Delta \nu_i^{a}$ remain consistent with linearity, one can then set constraints on potential new physics. 
Conversely, any significant deviation from linearity may signal either higher-order SM corrections or the existence of new, isotope-dependent forces coupling electrons and neutrons \cite{Berengut:2017zuo, Flambaum:2017onb, Frugiuele:2016rii}. 
    
Experimentally, we are now in a position where sub-kHz spectroscopy allowed to establish King NL in several systems, most notably
Yb\textsuperscript{+} \cite{Counts:2020aws,Hur:2022gof,Ono:2021ogd,Door:2024qqz} and Ca\textsuperscript{+} ions \cite{Rehbehn:2021zlr, Wilzewski:2024wap}.
There are many more systems where King NL could be measured in the future (see, e.g.,~\cite{Solaro:2020dxz}). 
This remarkable experimental progress raises the question of whether such high-precision data can also be leveraged to probe nuclear structure with unprecedented accuracy.
The central difficulty in describing and interpreting King NL is the limited theoretical control over non-perturbative SM contributions (see, for instance, ref.~\cite{Berengut:2025nxp} for a recent review). 
Understanding the SM contributions remains an active area of research, and it is not yet fully settled whether all relevant sources have been correctly identified. In the pioneering work of ref.~\cite{Counts:2020aws}, which first established the observation of King NL, the $\langle r^4 \rangle$ and $\langle r^2 \rangle^2$ contributions were, for instance, treated as related effects. Subsequent studies, however, recognized them as distinct contributions \cite{Hur:2022gof}, which received some support by the \emph{ab initio} nuclear-structure calculations presented in~\cite{Door:2024qqz}. The situation is further complicated by the fact that not all SM corrections generate independent sources of King NL. Some higher-order shifts in the electronic transition energy can be absorbed into common prefactors in the King linearity relation, thereby obscuring which effects genuinely contribute to the King NL.

What makes the problem more manageable, however, is that the physics involved is described by several well separated scales, so that one can construct the appropriate effective field theories (EFTs) and develop the corresponding counting of different contributions. 
In this paper we initiate such a systematic treatment of King NL using EFTs. The key observation is that there is a large hierarchy between typical nuclear,  $r_N \sim 10^{-15} \ \mathrm{m}$, and typical atomic length scale, $a\sim 10^{-10} \ \mathrm{m}$. Along with an expansion in the fine structure constant $\alpha$, this then allows to organize corrections to the atomic energy levels in terms of their scaling with $\epsilon=r_N/a\sim 10^{-5}$ and $\alpha\sim 10^{-2}$. Technically, this is done by constructing two EFTs, and performing a two-step matching. In the first step, we match the SM onto non-relativistic quantum electrodynamics (NRQED), where both the nucleus and the electrons are treated as non-relativistic objects described by second quantized fields. To simplify the discussion, we work in this first step in the infinite nuclear mass limit. In the second step, we match onto a first-quantized description, with interactions now described by the quantum-mechanical  (QM) potential between electrons and nucleus (sometimes also referred to as potential NRQED  or pNRQED  \cite{Pineda:1998kn}). This potential can then be applied systematically within QM perturbation theory.

The above EFT framework cleanly separates short-distance nuclear physics and long-distance atomic physics. It also provides a transparent classification of which operators contribute to King NL at a given level of precision.  While the approach is applicable more generally, in this paper we apply it to a hydrogen-like system with spin-0 nuclei. Even in this simple setting, we are able to gain better understanding of several SM contributions to the King NL. In particular, we show that the common treatment of the $\langle r^2\rangle^2$ term in the literature is incorrect: it arises only through second-order QM perturbation theory, requiring the explicit use of electronic Green's function, and not just the wave function of the state itself.  In addition, we derive, using field theory methods, the parametrically more important contribution of nuclear polarizability to King NL, demonstrating that it induces a long-range $1/r^4$ potential, and provide estimates of its impact on King NL. 
 
The paper is organized as follows. In section~\ref{sec:eft} we construct the scalar NRQED Lagrangian in the infinite nuclear mass limit and perform the first matching from the SM, identifying the relevant Wilson coefficients such as the charge radius squared $\langle r^2 \rangle$, $\langle r^4 \rangle$, and polarizabilities. In section~\ref{sec:qm} we match NRQED onto QM potentials (pNRQED), including  tree-level and loop-induced interactions, and highlight which operators contribute to genuine King NL. In section~\ref{sec:energy_levels} we calculate the higher order corrections to energy levels,  and clarify the origin of the $\langle r^2 \rangle^2$ term. Section~\ref{sec:pheno} discusses the nonlinearities in isotope-shift experiments for Ca and Yb, emphasizing the separation of SM contributions from possible new physics. We conclude in section~\ref{Summary}, while appendices contain further details on the material covered in the main text:  app. \ref{sec:app:matching} contains details on matching calculation of NRQED Wilson coefficients, app. \ref{Appendix:Box} on matching onto QM potentials, app. \ref{Appendix:UV_IR_divergencies} a discussion of UV and IR divergences, and app. \ref{sec:app:r22} further details on $\langle r^2\rangle $ induced corrections to the wave functions.

\section{NRQED for King Non-linearity}
\label{sec:eft}

We start with the first step in the matching procedure, matching the SM onto NRQED, where both the nucleus and the electron are treated as non-relativistic. We limit the discussion to hydrogen-like systems, composed of a single nucleus and a single electron, so that no electron-electron interactions are needed. We also limit ourselves to the case of spin-0 nuclei, which are then described by a scalar field $S$. This reduces the number of possible NRQED operators, simplifying the discussion, and compatible with systems where King NL was observed. In sec.~\ref{Summary}, we discuss how one can go beyond this approximation using our EFT framework. 

The matching of SM onto NRQED for the case of hydrogen, i.e., an EFT for a  non-relativistic spin-1/2 nucleus interacting with a non-relativistic electron is well known~\cite{Manohar:1997qy, Pineda:1998kn, Hill:2012rh, Gunawardana:2017zix}, and we can reuse many of the results. However, there is also an important difference: unlike for the case of a proton, we need to keep track separately of the scalings of different Wilson coefficients of NRQED operators with the inverse size of the nucleus, $1/r_N\sim A^{-1/3}/r_p$, and with its mass, $M\sim A m_p$. Here, $A$ is the mass number and is $A\gg 1$ in the systems we are interested, e.g., $A(\text{Yb})\simeq 170$ and  $A(\text{Ca})\simeq 40$. That is, while in the proton-electron system one can still treat the proton mass, $m_p\simeq 1$~GeV, and the inverse of proton radius, $1/r_p \simeq 1$ fm$^{-1} \simeq 0.2$~GeV 
\cite{ParticleDataGroup:2024cfk} as though they are the same scale, this is no longer true even for a ``lighter" nucleus such as Ca, with charge radius $1/r_N\simeq 1/(3.5 \text{ fm})\simeq 0.06$~GeV \cite{Angeli:2013epw},  and mass is $M\simeq 40$~GeV \cite{ParticleDataGroup:2024cfk}. 
This difference is even larger for Yb, for which 
$1/r_N\simeq 1/(5\,\text{fm})\simeq 0.04$~GeV \cite{Angeli:2013epw}, and $M\sim 160$~GeV \cite{ParticleDataGroup:2024cfk}. 

In the two-body problem of a single electron bound to a nucleus, we thus have the following hierarchy of scales
\beq
E_N\ll E_e\ll p\ll m_e\ll \Lambda_N \ll M, 
\eeq
where $\Lambda_N\equiv 1/r_N$ denotes the typical scale associated with the nucleus being an extended object, $m_e$ is the mass of the electron (which is almost equal to the reduced mass of the system), $p\sim Z\alpha m_e $ is the typical momentum of electron and/or nucleus, $E_e\sim p^2/m_e\sim Z^2 \alpha^2 m_e$ is the typical kinetic energy of the bound electron, and $E_N\sim p^2/M\sim  Z^2 \alpha^2 m_e^2/M$ the kinetic energy of a nucleus. Since $M$ is much larger than any of the other energy scales in the problem, we can safely take the $M\to \infty$ limit to the precision we are working when constructing the NRQED Lagrangian relevant for King NL calculations. Note that in this limit also $E_N\to 0$, further simplifying the problem.

The NRQED Lagrangian for a hydrogen-like system with spin-0 nucleus is composed out of three terms, 
\beq
{\cal L}_\text{NRQED}={\cal L}_{S}+{\cal L}_{S\psi_e}+{\cal L}_{\psi_e},
\eeq
with ${\cal L}_{S}$~$({\cal L}_{\psi_e})$ describing the interactions of a non-relativistic scalar nucleus field $S$ (non-relativistic electron field $\psi_e$) with the electromagnetic field, while  ${\cal L}_{S\psi_e}$ contains contact interaction terms between $S$ and $\psi_e$. In the $M\to \infty$ limit ${\cal L}_{S}$ is obtained as an expansion in $p/\Lambda_N$, ${\cal L}_{\psi_e}$ as an expansion in $p/m_e$, while ${\cal L}_{S\psi_e}$ involves both expansions. In the construction of the higher dimension operators the non-relativistic fields and derivatives scale as  (after the $M\to \infty$ limit was already taken, note that we do not distinguish between soft and ultrasoft modes \cite{Luke:1999kz})
\beq
S\sim p^{3/2}, \qquad \psi_e \sim p^{3/2}, \qquad \partial_i \sim p, \qquad \partial_t \sim p^2/m_e,
\eeq
while for the electromagnetic vector potential, the electric and magnetic fields, we have (see \cite{Lepage:1992tx})
\beq \label{EM_Power_Counting}
A_0 \sim p^2/m_e, \qquad A_i \sim p, \qquad E_i\sim p^2, \qquad B_i \sim p^3/m_e.
\eeq
We will need the explicit forms only for ${\cal L}_{S}$ and ${\cal L}_{S \psi_e}$, given in eqs. \eqref{L_S} and \eqref{contact_interactions} below, while ${\cal L}_{\psi_e}$ (up to  dimension 8) can be found in \cite{Hill:2012rh,Gunawardana:2017zix}.

Let us first derive ${\cal L}_{S}$: at the present level of experimental precision, it suffices to include operators up to dimension 8. We also only need interactions that are bilinear in $S$. The complete basis of such operators was explicitly constructed in the context of non-relativistic quantum chromodynamics (NRQCD) in Ref.~\cite{Gunawardana:2017zix} (see also \cite{Kobach:2017xkw}), and can also be inferred from \cite{Hill:2012rh} by discarding the spin-dependent terms. Importantly, Lorentz invariance relates some of the higher-dimension Wilson coefficients to the lower ones \cite{Luke:1992cs,Heinonen:2012km,Hill:2012rh} (this is often referred to as the reparametrization invariance (RPI) when working to leading order).
In the infinite-mass limit these relations imply that many of the higher-dimension operators vanish, reducing the number of operators we need to consider. We will demonstrate this explicitly below. 

To construct the form of ${\cal L}_{S}$ we define first (see also app.~\ref{One_photon_Wilson})
\beq\label{minimal_coupling}
D_t=\partial_t+ieZA^0, \quad \bm D=\bm\nabla-ieZ\bm A,
\eeq
where $e$ is the electromagnetic coupling constant and $Z$ is the total electric charge of the nucleus. In eq.~\eqref{minimal_coupling} we follow the convention of ref.~\cite{Manohar:1997qy}. At dimension 4, ${\cal L}_{S}$ contains  a single operator, ${\cal L}_{S}\supset S^\dagger i D_t S$.
At dimension 5, there is similarly only a single spin-independent operator,  ${\cal L}_{S}\supset c_2 S^\dagger \bm D^2 S/2M$, where Lorentz invariance fixes $c_2=1$. In the $M\to\infty$ limit this term vanishes.  At dimension 6 there is also a single spin-independent operator, which, however, does not vanish in the $M\to \infty$ limit. Conventionally, in the cases where $M$ and $\Lambda_N$ are not parametrically different, this is written as ${\cal L}_{S}\supset c_D S^\dagger e (\bm\nabla\cdot \bm E) S/8M^2$. However $c_D/M^2$ does not scale with the mass of the nucleus, but rather with the size of the nucleus. We make this explicit, by setting $c_D/8M^2={\coneprime}/{\Lambda_N^2}$, where ${\coneprime}/{\Lambda_N^2}$ is proportional to $\langle r^2\rangle$, the squared charge radius of the nucleus (see app.~\ref{One_photon_Wilson}). At dimension 7 there are four spin-independent operators: $c_4S^\dagger \bm D^4 S/8M^3$, 
$i c_M g\, S^\dagger \{ \bm D^i , [\bm\partial \times \bm B]^i \} S/16M^3$,
$c_{A1} g^2\, S^\dagger (\bm B^2 - \bm E^2) S/8M^3$, $c_{A2} g^2\, S^\dagger \bm E^2 S/16M^3$ \cite{Manohar:1997qy}.
Lorentz invariance implies $c_4=1$ and $c_M=c_D$ \cite{Hill:2011be}, so the first two vanish as $M\to\infty$. The Wilson coefficients of last two are, however, related to the electric ($\alpha_E$) and magnetic ($\beta_M$) polarizabilities, which scale as $\sim r_N^3$ and do not vanish in the $M\to \infty$ limit, see \cite{Hill:2012rh}.
Finally, at dimension 8 there are four spin-independent operators: $c_{X1}S^\dagger g [ \bm D^2 , \bm D\cdot \bm E + \bm E\cdot \bm D ] S/M^4$,  $c_{X2}S^\dagger g \{ \bm D^2 , (\bm\nabla \cdot \bm E) \} S /M^4$, $c_{X3}S^\dagger g [\bm\nabla^2 (\bm\nabla\cdot \bm E)] S/M^4$,
$ic_{X4}S^\dagger g^2 \{ \bm D^i , [\bm E \times \bm B]^i \} S/M^4$ \cite{Hill:2012rh}.
Lorentz invariance relates $c_{X1}$, $c_{X2}$, and $c_{X4}$ to lower-dimensional coefficients \cite{Hill:2012rh}. Therefore, only one operator, $c_{X3}\, S^\dagger g [\bm\nabla^2(\bm\nabla\cdot \bm E)] S/M^4$, is non-vanishing in the $M\to \infty$ limit, namely,  $c_{X3}/M^4=\cfourprime/{\Lambda_N^4}$, where $\cfourprime/{\Lambda_N^4}$ is proportional to $\langle r^4\rangle$.

Collecting the above results, the scalar NRQED Lagrangian in the $M\to\infty$ limit takes the simple form  
\begin{equation}\label{L_S:init}
{\cal L}_{S} = S^\dagger\left\{ i D_t 
+ \frac{\coneprime}{\Lambda_N^2}\,(\bm\nabla\!\cdot\!\bm E)
+ \frac{\ctwoprime}{\Lambda_N^3}\,\bm E^2 
+ \frac{\cthreeprime}{\Lambda_N^3}\,\bm B^2
+ \frac{\cfourprime}{\Lambda_N^4}\,\bm\nabla^2(\bm\nabla\!\cdot\!\bm E)
\right\}S \,,
\end{equation}
where we denoted explicitly the expected scaling with $\Lambda_N$, and kept terms up to dimension 8. Here, the $\coneprime$ term gives the charge radius of the nucleus, $\langle r^2\rangle$, the $\ctwoprime$ and $\cthreeprime$ are related to the electric and magnetic polarizabilities of the nucleus, respectively, while the $\cfourprime$ is related to $\langle r^4 \rangle$ (where, intuitively, as for the charge radius, the expectation value is over the electric charge distribution inside the nucleus; see app.~\ref{One_photon_Wilson} for the exact definition). Including all the prefactors, we have\footnote{Eq.~\eqref{L_S} uses Gaussian units, including for the definitions of $\alpha_E S^\dagger E^2S/2$ and $\beta_M S^\dagger B^2S/2$ operators,  see \cite{Schumacher:2005an}. This should be compared with the definitions that use Heaviside units, e.g., in ref.~\cite{Hill:2012rh}, which in the $M\to \infty$ limit  uses $ e^2\alpha_E S^\dagger E^2S/(2\alpha)=2\pi \alpha_ES^\dagger E^2S$ and $e^2\beta_M S^\dagger B^2S/(2\alpha)=2\pi\beta_M S^\dagger B^2 S$ as the definitions for the same operators, see \cite{Schumacher:2005an}. The difference is an overall factor of $4\pi$ \cite{Schumacher:2005an}, which is already taken into account in the expressions below. That is, the model estimates for $\alpha_E$ below are in the convention used in \eqref{L_S}.}
\begin{equation}
\label{L_S}
{\cal L}_{S} = S^\dagger\left\{ i D_t 
+ eZ\,\frac{\langle r^2\rangle}{6}\,(\bm\nabla\!\cdot\!\bm E)
+ \frac{\alpha_E}{2}\,\bm E^2 
+ \frac{\beta_M}{2}\,\bm B^2
+ eZ\,\frac{\langle r^4\rangle}{120}\,\bm\nabla^2(\bm\nabla\!\cdot\!\bm E)
\right\}S \, .
\end{equation}
That is, up to dimension~8 in the infinite-mass limit, the interaction of a scalar nucleus with photons is fully characterized by just four Wilson coefficients: $\langle r^2\rangle$, $\langle r^4\rangle$, and the electric and magnetic polarizabilities $\alpha_E,\beta_M$. 

We include the $\beta_M\, S^\dagger\bm B^2S/2$ operator, although according to eq. (\ref{EM_Power_Counting}) it is power suppressed. We will use it in the calculation of the polarizabilities' contribution to the potential. As we will see below, the magnetic polarizability will only contribute to the short-distance part of the potential arising from the interaction with the relativistic electrons. In accordance with  eq. (\ref{EM_Power_Counting}) and at the order we are working, it does not contribute to the long-range potential arsing from  the interaction with non-relativistic electrons, see section \ref{sec:polarizability}. 

The value of the electric polarizability can be estimated from the hydrodynamic model: for heavy nuclei, it is proportional to atomic mass number as $A^{5/3}$ \cite{Migdal:1945, Levinger:1957zz}. In this model it is given as~\cite{Flambaum:2017onb}

\beq
\begin{split}\label{eq: elect_pol_1}
\alpha_E &= \left(0.8 + 2.8 A^{-1/3} \right) \frac{e^2 R^2  A}{40 \, a_{\rm sym}}=\left(0.8 + 2.8 A^{-1/3} \right) 2.6\times10^{-2}\, A^{5/3}\ \rm fm^3,
\end{split}
\eeq
where $e^2=4\pi\alpha$, $a_{\rm sym} = 23\,$MeV\,(0.12 $\text{fm}^{-1}$) is the nuclear symmetry energy and $R$ is the nuclear radius. For $R = r_0 A^{1/3}$, where $r_0 = 1.15$ fm \cite{Flambaum:2017onb}, the estimated values of electric polarizability from eq.~\eqref{eq: elect_pol_1} for Ca ($40 \leq A \leq 46$) and Yb ($168 \leq A \leq 174$) isotopes are respectively in the range 
$\alpha_E\in [19, 24]\,\rm fm^3$ and $\alpha_E\in[170, 180] \, \rm fm^3$.
Alternatively, using data from global analysis of photo-absorption cross-section measurements \cite{Orce:2023qtm}, the magnitude of $\alpha_E$ can be expressed analytically as~\cite{Liu:2025ows}, 

\begin{align}\label{eq:elect_pol_2}
    \alpha_E = 32 \pi \frac{(A/132)^2}{(A/132)^{1/3}-0.31} \, \rm fm^3.
\end{align}
Using eq.~\eqref{eq:elect_pol_2}, the values for $\alpha_E$ for Ca and Yb isotopes are, respectively, in the range
$[26, 31] \, \rm fm^3$ and $[210, 220] \, \rm fm^3$. Comparison between the two estimates shows that the estimated values of $\alpha_E$ carry about $20\%$ modeling uncertainty. Perhaps even more importantly, the two predictions for $\alpha_E$ have different dependence on $A$, and thus predict different trends for $\alpha_E$ variation between different isotopes of a given element. 
Note, that we will not need the value of the magnetic polarizability $\beta_M$ in our analysis. 

Next, let us turn to the interactions of non-relativistic electrons, given by ${\cal L}_{S\psi_e}$ and ${\cal L}_{\psi_e}$. The choice of performing the non-relativistic expansion also for electrons is motivated by the fact that in the QM analysis of ref.~\cite{Friar:1978wv}, relativistic corrections do not generate new sources of King NL. In particular, the relativistic corrections can be absorbed into the field shift, see section \ref{sec:friar}. The only notable exception is the case of polarizability, where matching onto QM potentials requires going beyond the non-relativistic approximation. These contributions, absent in ref.~\cite{Friar:1978wv}, are treated in sec.~\ref{sec:polarizability}.

The bilinear Lagrangian up to dimension 8 of heavy fermions and non-relativistic electrons can be found in ref.~\cite{Hill:2012rh}, from which we can infer the Lagrangian that contains two non-relativistic electron fields and two non-relativistic scalar fields up to dimension 8.
We follow the notation of ref.~\cite{Hill:2012rh} for the Wilson coefficients, 
except that we now include the expected scaling with  $\Lambda_N$. Denoting the non-relativistic electron field by $\psi_e$ we have, for the terms that are expected not to vanish in the $M\to \infty$ limit,

\beq
 \begin{split}\label{contact_interactions} 
 {\cal L}_{S\psi_e} =&\dfrac{d_2 m_e}{\Lambda_N^3}S^\dagger S\psi_e^\dagger\psi_e+\dfrac{d_3}{\Lambda_N^4}S^\dagger D_+^i S\psi_e^\dagger D_+^i \psi_e
 +\dfrac{d_6}{m_e \Lambda_N^3}S^\dagger  S\psi_e^\dagger (\bm{D}^2 + \overleftarrow{\bm{D}}^2 )\psi_e
 \\
 &+\dfrac{g d_{10}}{\Lambda_N^4} S^\dagger S \ \psi_e^\dagger \bm{\sigma}\cdot\bm{B} \psi_e + \dfrac{i d_{11}}{m_e \Lambda_N^3} \epsilon^{ijk} S^\dagger D_+^k S \ \psi_e^\dagger \sigma^i D_-^j \psi_e,
 \end{split}
 \eeq
 where  $D_\pm =D\pm \overleftarrow{D}$, while the terms suppressed by powers of $1/M$ are
 \beq
 \begin{split}\label{contact_interactions_alternative:1/M}
 {\cal L}_{S\psi_e}^{1/M} =&\dfrac{d_4}{M \Lambda_N^3}S^\dagger D_-^i S\psi_e^\dagger D_-^i \psi_e
 +\dfrac{m_e d_5}{M^2 \Lambda_N^3}S^\dagger (\bm{D}^2 + \overleftarrow{\bm{D}}^2 ) S\psi_e^\dagger \psi_e
 \\
&+ \dfrac{i d_{12}}{M \Lambda_N^3} \epsilon^{ijk} S^\dagger  D_-^k S \ \psi_e^\dagger \sigma^i  D_+^j \psi_e 
\,.
 \end{split}
 \eeq
 
As we will show in section \ref{sec:energy_levels} below, out of the operators in eqs.~\eqref{contact_interactions} only the operator with the Wilson coefficient $d_6$ induces a new contribution to King NL.  Nevertheless, to compare with previous literature \cite{Friar:1978wv}, we need the value of the Wilson coefficient for the lowest-dimension contact interaction, $d_2$. This term arises at one loop, from a two-photon exchange between the electron and the nucleus, so that $d_2\sim {\mathcal O}(Z^2\alpha^2)$.\footnote{The tree level contribution to the $S^\dagger S\psi_e^\dagger\psi_e$ operator, from a single photon exchange, is included as the $\langle r^2\rangle$ term in \eqref{L_S}. It is point-like since the $1/q^2$ from the photon propagator is canceled by the linear $q^2$ dependence in the charge radius part of the electromagnetic form factor for $S$; see also eq.~\eqref{eq:r2:potential} below.}
 The additional $m_e/\Lambda_N$ suppression in the definition of $d_2$ is obtained from the matching in appendix \ref{Two_photon_Wilson}. We find that  
 \beq
 \label{eq:d_2:matching:result}
 \frac{d_2}{\Lambda_N^3}=\frac{\pi}{3} \left(Z\alpha\right)^2\langle r^3\rangle_{(2)} +\frac{d_2^\text{\,inel.}}{\Lambda_N^3}+\frac{d_2^\text{1/M}}{m_e^2 M}.
\eeq
 where $\langle r^3\rangle_{(2)}\sim {\mathcal O}(1/\Lambda_N^3)$ is the well-known Friar radius of the nucleus \cite{Friar:1978wv, Eides:2000xc}. The size of the inelastic contribution $d_2^\text{\,inel.}$ is currently not known for heavy nuclei, but was, for instance, found to be ${\mathcal O}(50\%)$ of the elastic one for proton structure corrections to the Lamb shift in muonic hydrogen \cite{Carlson:2011zd}. In \eqref{eq:d_2:matching:result} we also included the $1/M$ suppressed contribution, which for a point-like spin 1/2 particle in the $\overline{\mbox{MS}}$ scheme is given by $d_2^{1/M}= Z^2\alpha^2 \big(\log(m_e^2/\mu^2)+1/3\big)$~\cite{Pineda:1998kn}. Since 
 we work in the $M\to\infty$ limit in this paper, $d_2^\text{1/M}/{m_e^2 M}$ is formally power suppressed. Numerically, since $\Lambda_N^3\sim 20(2) m_e^2 M$ 
 for Ca (Yb), it is, however, expected to be the most important term in $d_2$  for nuclei we consider. Since we have a composite scalar, and not a point-like spin 1/2 particle, our analysis needs to be extended to order $1/M$ to fully determine the magnitude of $d_2^\text{1/M}$. We leave that for a future study.

Since they are not required for our purposes, we will not explicitly calculate the other Wilson coefficients in eqs.~\eqref{contact_interactions} and \eqref{contact_interactions_alternative:1/M}. The assigned scalings with $\Lambda_N$, $m_e$ and $M$ in the definitions of the operators are partially motivated by Lorentz invariance relations, a.k.a. the reparametrization (RPI) relations \cite{Luke:1992cs}, which require $d_2/4=d_4+d_5$, $d_5=d_6$, and $8(d_{11}+d_{12})=-d_2$ \cite{Hill:2012rh}. Since there are only three RPI relations among six Wilson coefficients, the higher-dimensional coefficients $d_4, d_5, d_6, d_{11},$ and $d_{12}$ cannot be uniquely expressed in terms of $d_2$ alone. In particular, while the assigned scalings for these operators in 
eqs.~\eqref{contact_interactions}~and~\eqref{contact_interactions_alternative:1/M} are consistent with the RPI relations, some of these Wilson coefficients could contain additional factors of $m_e/\Lambda_N$ and/or $\Lambda_N/M$. Furthermore, $d_3$ and $d_{10}$ do not enter the RPI relations at all, so that for these we simply used the naive dimensional scaling with $1/\Lambda_N$ in the definitions in eq.~\eqref{contact_interactions}.  The $1/M$ suppressed operators in eq.~\eqref{contact_interactions_alternative:1/M} are not expected to be important numerically, and we thus do not include them in our study from now on.

\section{Quantum mechanical potentials from NRQED}
\label{sec:qm}

In this section~we match the NRQED electron--scalar-nucleus interactions, eqs.~\eqref{L_S} and \eqref{contact_interactions}, onto QM potentials, working to leading power in $1/m_e$ and in the $M\to \infty$ limit. The resulting potential receives several different contributions
\beq
V(\bm r)= V_\text{1-photon}(\bm r)+V_\text{contact}(\bm r)+V_\text{pol.}(\bm r),
\eeq
and will then be used in QM perturbation theory to calculate the atomic energy levels and hence the isotope shifts. In this section we first perform the matching at tree level in NRQED: $V_\text{1-photon}(\bm r)$ from single photon exchange in sec. \ref{sec:Tree}, and $V_\text{contact}(\bm r)$ from contact terms in sec.~\ref{sec:contact}. The one-loop two-photon exchange calculation involving electric polarizability operator, which gives rise to $V_\text{pol.}(\bm r)$, is performed in sec.~\ref{sec:polarizability}. Other one-loop contributions, which are not relevant for King NL, are discussed in sec. \ref{sec:one-loop:other}.

\subsection{Tree level matching: single photon exchange} 
\label{sec:Tree}
\begin{figure}
    \centering
    \includegraphics[width=1\linewidth]{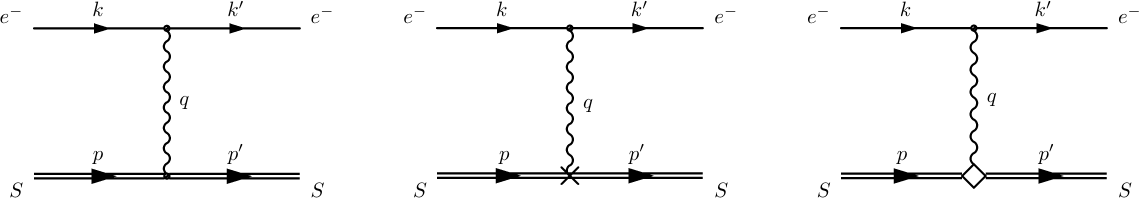}
    \caption{NRQED
    Feynman diagrams for one photon exchange contributing to the QM potential $V(r)$. The dot, cross, and diamond vertices represent the minimal coupling (from $D_t$), and from $\langle r^2 \rangle$, and $\langle r^4 \rangle$ terms, respectively.}
    \label{fig:one_photon_exchange}
\end{figure}

The QM potential that arises from a single photon exchange between a non-relativistic electron and a scalar nucleus receives three distinct contributions to the order we are working (see fig.~\ref{fig:one_photon_exchange}). These are due to the three distinct couplings of the scalar nucleus with a single photon, contained in eq.~\eqref{L_S}: the $S^\dagger i D_t S$ term, and the terms proportional to 
 charge radius squared, $\langle r^2\rangle$, and the fourth moment $\langle r^4\rangle$. The corresponding Feynman rules for an emission of $A^0$ are (see also~\cite{Kinoshita:1995mt,Dye:2018rgg}),
\beq
-iZe, 
\qquad \frac{iZe}{6}\,\langle r^2\rangle\,\bm{q}^{\,2},
\qquad -\frac{iZe}{120}\,\langle r^4\rangle\,(\bm{q}^{\,2})^2.
\eeq
In non-Coulomb gauges there are also contributions from the $A^i$ exchanges.  These are $1/m_e$  suppressed since $A^i$ couplings start at order $1/m_e$ in the  NRQED Lagrangian for the electron, ${\cal L}_{\psi_e}$, and can be neglected to the order we work. Thus from ${\cal L}_{\psi_e}$, we only keep the ${\mathcal O}(1/m_e^0)$ term, ${\cal L}_{\psi_e}= \psi_e^\dagger i D_t \psi_e +\cdots = -Z_ee\,\psi_e^\dagger A^0 \psi_e +\cdots$, where $Z_e=-1$.  

Equating each diagram in fig.~\ref{fig:one_photon_exchange} to $i\mathcal M$, the potential in momentum space is given by $\tilde V(\bm q)=-\mathcal M$ \cite{Peskin:1995ev}, while the potential in the position space is given by the Fourier transform
\begin{equation}
V(\bm r) \;=\;\int \frac{d^3 q}{(2\pi)^3}\, e^{i \bm q\cdot \bm r}\,\tilde V(\bm q).
\end{equation}
Summing the three contributions gives the tree level potential due to a single photon exchange, up to dimension 8, 
\beq
\label{eq:V:1-photon}
V_\text{1-photon}(\bm r)=V_Z(r)+V_{\langle r^2\rangle}(r)+V_{\langle r^4\rangle}(r).
\eeq
For the three contributions to the potential we find, 
\begin{align}
 \label{eq:Z:potential}
    i\mathcal{M}_Z &= \frac{c}{q^2}
    \qquad\quad~~\, \Rightarrow \quad
   \, \tilde{V}_Z(\bm{q}) = \frac{-Ze^2}{\bm{q}^{\,2}}
    \qquad\quad~\, \Rightarrow \quad
    V_Z(r) = -\frac{Z\alpha}{r},
    \\
    \label{eq:r2:potential}
    i\mathcal{M}_{\langle r^2\rangle} &= -\frac{c\langle r^2\rangle}{6q^2} {\bm q}^{2}
    ~~\, \Rightarrow ~\,
    \tilde{V}_{\langle r^2\rangle}(\bm{q}) = \frac{Z e^2}{6}\langle r^2\rangle
    \quad~~~~\,\Rightarrow\,
    V_{\langle r^2\rangle}(\bm r) = \frac{4\pi Z\alpha}{6}\langle r^2\rangle\,\delta^3(\bm r),
    \\
    i\mathcal{M}_{\langle r^4\rangle} &= \frac{c \langle r^4\rangle}{120q^2}(\bm q^{2})^2 
    \, \Rightarrow \,
    \tilde{V}_{\langle r^4\rangle}(\bm{q}) = -\frac{Ze^2}{120}\langle r^4\rangle{\bm q}^{2}
    ~~\,\, \Rightarrow \,
    V_{\langle r^4\rangle}(\bm r) = \frac{4\pi Z\alpha}{120}\,\langle r^4\rangle\,\nabla^2\left[\delta^3(\bm r)\right],
\end{align}
where in writing down the amplitudes we defined $c \equiv (-iZ_ee)(-i g_{00})\,\chi^\dagger_{s'}\chi_s(-iZe)=-ie^2\delta_{s's} Z$, with $\chi_s$ the two-component spinor for the electron. For $\tilde V(\bm q)$  we used the approximation $q^2\simeq -\bm q^{\,2}$. In the potentials we do not display $\delta_{s's}$, and write $e^2=4\pi \alpha$. We also used that the Fourier transform of ${\bm q}^{2}$ is $\nabla^2\left[\delta^3(\bm r)\right]$ \cite{Gel'fand}.

\subsection{Tree level matching: contact interactions}
\label{sec:contact}
Next, let us obtain the contributions to  the potential from the operators with four fields in eq.~\eqref{contact_interactions},\footnote{As we will see below, the derivatives acting on nuclear field $S$ in these operators can be moved by integration by parts to act on electron fields. This is not the case for contact operators in eq.~\eqref{contact_interactions_alternative:1/M}, and thus their expectation values necessarily involve the highly suppressed nuclear momenta and can be ignored.} 
\beq
V_\text{contact}(r)=\sum_i V_{i}(r),
\eeq
where the sum is over the operators with Wilson coefficients $i=d_2, d_3, d_6, d_{10}, d_{11}$.

The single dimension 6 four-field operator,
\begin{equation}
{\cal L}_{d_2}=\frac{d_2 m_e}{\Lambda_N^3}\,S^\dagger S\,\psi_e^\dagger\psi_e\,,
\end{equation}
produces a momentum-independent amplitude and hence a local potential,
\begin{equation}
{i\mathcal M}_{d_2} = i\frac{d_2 m_e}{\Lambda_N^3}\quad\Rightarrow\quad \tilde V_{d_2}(\bm q) = -\frac{d_2 m_e}{\Lambda_N^3}\quad\Rightarrow \quad V_{d_2}(r) = -\frac{d_2 m_e}{\Lambda_N^3}\,\delta^3(\bm r)\, .
\end{equation}
Note that this contribution to the potential has the same $r$ dependence as $V_{\langle r^2\rangle}(r)$ in eq.~\eqref{eq:r2:potential} even though $V_{d_2}(r) $ is suppressed by ${\mathcal O}(Z \alpha m_e/\Lambda_N)$ (see also the discussion below eq. \eqref{contact_interactions_alternative:1/M} and in app.~\ref{Two_photon_Wilson}).

Next, we consider the dimension-8 operators, eq.~\eqref{contact_interactions}. The operator with the Wilson coefficient $d_{10}$ will only generate a shift to the energy level when coupled to an external magnetic field. Therefore we do not include it in the potentials below.    
Two of the operators in eq.~\eqref{contact_interactions} include covariant derivatives on $S$. Since these operators are already power suppressed, we can ignore, to the order we are working, the coupling to photons in the covariant derivatives and use integration by parts to move the regular derivative to act on the electron fields:
\begin{align}
    {\cal L}_{d_3}&=-\frac{d_3}{\Lambda_N^4}\big(S^\dagger S\big) \left[2\left(\bm{\nabla}\psi_e^\dagger\cdot   \bm{\nabla}\psi_e\right)+\psi_e^\dagger (\overleftarrow{\nabla}^2+\nabla^2) \psi_e\right]+{\cal O}(e) \,, \nonumber\\
    &=-\frac{d_3}{\Lambda_N^4}\big(S^\dagger S\big) \nabla^2\left[\psi_e^\dagger\psi_e\right]+{\cal O}(e) \,, \\
 {\cal L}_{d_{11}}&=\dfrac{2id_{11}}{m_e \Lambda_N^3}\big(S^\dagger S\big) \left[\epsilon^{ijk}(\partial_j \psi_e)^\dagger \sigma^i (\partial_k \psi_e)\right] +{\cal O}(e)\,, 
\end{align}
while for the $d_6$ term no changes are needed compared to the definition of the operator in eq.~\eqref{contact_interactions},
\begin{align}
   {\cal L}_{d_6}&=\frac{d_6}{m_e\Lambda_N^3}\big(S^\dagger S\big) \left[\psi_e^\dagger (\overleftarrow{\nabla}^2+\nabla^2) \psi_e\right]+{\cal O}(e) \,.
\end{align}
To obtain the corresponding QM potentials for the operators with the Wilson coefficients $d_{3,6,11}$, it is more straightforward to not use the Fourier transform, but rather to require that ${\cal M}=- \int d^3 \bm r \langle \phi| V(\bm r)| \phi \rangle$, where the expectation is over the electronic wave function, knowing that the potential will be proportional to $\delta^3(\bm r)$ with appropriately placed derivatives.\footnote{Alternatively, one can follow the approach of Ref.~\cite{Pineda:1998kn}, where, neglecting spin, the potential $\tilde V$ depends on two distinct variables: $\bm q$, the Fourier transform variable of $\bm r$, and $\bm p=-i\bm{\nabla}_{\bm r}$. Both approaches lead to the same position-space potentials. An analogous procedure to Ref.~\cite{Pineda:1998kn} was applied in the derivation of the spin-orbit charge density in Refs. \cite{Ong:2010gf,Caputo:2024doz}.} 
This gives
\begin{align}
V_{d_3}(\bm r)&=\frac{d_3}{\Lambda_N^4}\nabla^2\left[\delta^3(\bm r)\right] , 
\\
V_{d_6}(\bm r)&=-\frac{d_6}{m_e \Lambda_N^3}\big(\overleftarrow{\nabla}^2 \delta^3(\bm r)+\delta^3(\bm r) \nabla^2\big),
\\
 V_{d_{11}}(\bm r)&=-\frac{2id_{11}}{m_e \Lambda_N^3}\big[\bm{\sigma}\cdot\big(\overleftarrow{\bm{\nabla}}\delta^3(\bm r)\ \bm{\times}  \delta^3(\bm r)\bm{\nabla}\big)\big],
\end{align}
where we were careful to place the derivatives acting on the final state electronic wave function to the left of the delta functions. Explicitly, the expectation values are 
\begin{align}
\langle V_{d_{3}}\rangle_\phi
&=
\frac{d_3}{\Lambda_N^4}\,\Big\{
\left[\nabla^2\phi^*(0)\right]\phi(0)+2\,\bm\nabla\phi^*(0)\cdot \bm\nabla\phi(0)+\phi^*(0)\left[\nabla^2\phi(0)\right]\Big\},
\\
\langle V_{d_{6}}\rangle_\phi
&=-\frac{d_6}{m_e \Lambda_N^3}\, \Big\{
\big[\nabla^2\phi^*(\bm 0)\big]\phi(\bm 0)+
\phi^*(\bm 0) \nabla^2\phi(\bm 0)
\Big\} ,
\\
\langle V_{d_{11}}\rangle_\phi
&=-\frac{2id_{11}}{m_e \Lambda_N^3}\, 
\left\{\epsilon^{ijk}\big[\partial_j \phi^\dagger (0)\big] \sigma^i \big[\partial_k \phi(0)\big]\right\}\,,
\end{align}
where on the l.h.s. we introduced a short-hand notation
 \beq
 \label{eq:V:expect}
  \langle V_{d_{i}}\rangle_\phi \equiv \int d^3 \bm r  \langle\phi|V_{d_{i}}(\bm r)|\phi\rangle. 
  \eeq
The above contact operators thus probe the curvature of the electronic wave function at the origin. 

Note that the contributions to the energy levels from $V_{d_3}$ and $V_{d_6}$ are suppressed by, respectively,  ${\mathcal O}(k^2/m_e \Lambda_N)\sim {\mathcal O}(Z^2 \alpha^2 m_e/\Lambda_N  )$ and by ${\mathcal O}(k^2/m_e^2)\sim {\mathcal O}(Z^2 \alpha^2)$, compared to $V_{d_2}$. Here, both $d_2$ and $d_{3,6}$ are expected to first arise from one-loop two photon exchange contributions when matching onto NRQED, and thus we expect $d_2\sim d_{3,6} \sim {\mathcal O}(Z^2 \alpha^2)$. The effect of the $V_{d_{11}}$ potential, which involves the electron spin, is to split the $p$-level according to the electron spin 
\begin{eqnarray}
  \Delta E_{d_{11}}  = -\frac{3d_{11}}{\pi m_e\Lambda_N^3}\left[\dfrac{dR_{n1}}{dr}(0)\right]^2 m_\ell \,m_s, 
\end{eqnarray}
where $m_s=\pm1/2$ and we used the standard definition $\phi_{n1 m}(\bm{r})=R_{n1}(r)\,Y_{1m}(\theta,\varphi)$.
For $s$-level (where the triple product is zero) and $d$-levels (or higher), these contributions vanish, since for central potentials $\bm \nabla \phi(\bm 0)=0$ for $s$- and $d$-orbitals (or higher).

\subsection{One-loop matching: polarizability} \label{sec:polarizability}

\begin{figure}[t]
    \centering  \includegraphics[width=0.45\linewidth]{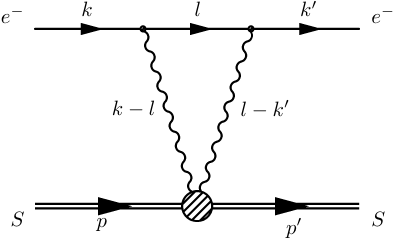}
    \caption{\label{fig:Feyn_Diag_Pol} Contribution from electric and magnetic polarizabilities (represented as a shaded circle) to the energy shift in a hydrogen-like atom.}
\end{figure}

In the $M\to \infty$ limit there are only two spin-independent dimension-7 operators --- the electric and magnetic polarizability operators in eq.~\eqref{L_S},
\begin{equation}
\label{eq:L_S:polarizab}
{\cal L}_{S} \supset S^\dagger\!\left(\frac{\alpha_E}{2}\,\bm E^{\,2}+\frac{\beta_M}{2}\,\bm B^{\,2}\right)\!S.
\end{equation}
To calculate the potential generated by these operators we use a shortcut: we calculate the diagram in fig.~\ref{fig:Feyn_Diag_Pol} using an EFT with a relativistic electron and a nonrelativistic nucleus, and then split the amplitude into a short-distance part (giving a contribution to the $d_2$ Wilson coefficient in eq.~\eqref{eq:d_2:matching:result}) and the leading non-local contribution, ${\cal M}_\text{pol.} (\bm{q}^2)$.\footnote{The latter could also be obtained by performing a calculation in NRQED, i.e., from a diagram in fig. \ref{fig:Feyn_Diag_Pol} with the electron line now denoting a non-relativistic field.} 

In covariant notation, eq.~\eqref{eq:L_S:polarizab} is written as
\begin{equation}
{\cal L}_{S} \supset S^\dagger\!\left[\frac{\beta_M}{4}\,F^{\mu\nu}F_{\mu\nu}
-\frac{(\alpha_E+\beta_M)}{2}\,F^{\mu\alpha}F^{\nu}{}_{\alpha}\,v_\mu v_\nu\right]\!S,
\end{equation}
where $F_{\mu\nu}$ is  the electromagnetic field-strength tensor, and  $v^\mu=(1,0,0,0)$ the heavy-nucleus four-velocity label, taken for simplicity to coincide with the lab frame. The corresponding Feynman rule is 
\beq
\begin{split}\label{eq:pol_feyn}
T_{\rho\sigma}
&=  
i\,\beta_M\!\left(g_{\rho\sigma}\,q_1\!\cdot\! q_2 - q_{1\sigma}q_{2\rho}\right)
- i\,(\alpha_E+\beta_M)\!\left[g_{\rho\sigma}(v\!\cdot\! q_1)(v\!\cdot\! q_2) \right. \\
&\hspace{7.1em}\left. -
q_{1\sigma}v_\rho (v\cdot q_2)-q_{2\rho}v_\sigma(v\cdot q_1)
+ (q_1\!\cdot\! q_2)\,v_\rho v_\sigma\right],
\end{split}
\eeq
where $v \cdot q_{1,2}=q_{1,2}^0$. In the $q_1=q_2$ limit this agrees with ref.~\cite{Szafron:2019tho}. The one-loop integral expression for the amplitude is then 
\begin{equation}
i\,\mathcal{M}_{\alpha_E, \beta_M}
= i e^2 \!\int\!\frac{d^d l}{(2\pi)^d}\;
\frac{\bar u(k')\,\gamma^\rho(\slashed{l}+m_e)\gamma^\sigma\,u(k)}{[(k-l)^2+i0][(l-k')^2+i0]\,[l^2-m_e^2+i0]}\;
T_{\rho\sigma},
\label{eq:Mpol}
\end{equation}
with $d=4-2\epsilon$, and $k,k'$ the four-momenta of incoming and outgoing electrons. We evaluate eq.~\eqref{eq:Mpol} in dimensional regularization using \texttt{FeynCalc} to perform the Dirac algebra and Passarino-Veltman reduction to a basis of scalar integrals $A_0$, $B_0$, $C_0$ \cite{Mertig:1990an,Shtabovenko:2016sxi,Shtabovenko:2020gxv}. Taking the nonrelativistic limit, $\bar u(k^\prime) u(k) \simeq \chi^\dagger_{s^\prime}\chi_s+{\mathcal O}(1/m_e^2)$, and setting
$k\cdot v =k^\prime \cdot v =m_e$ and $k\cdot k^\prime = m_e^2 - q^2/2$, we obtain from eq.~\eqref{eq:Mpol} the amplitude in the momentum space, 
\beq
\begin{split}
\label{eq:fin_amp}
    \mathcal{M}_{\alpha_E, \beta_M} \approx \frac{\alpha}{8\pi} \bigg\{\big(5\,\alpha_E-\beta_M\big) m_e\bigg[\frac{1}{\epsilon}-\gamma_E&+\log\Big(\frac{4\,\pi\,\mu^2}{m_e^2}\Big)\bigg]+8\alpha_Em_e
    \\
     &+ \pi^2 \,i\,\alpha_E  \sqrt{-\bm{q}^{2}-i0}\bigg\} + \mathcal{O}\Big(\frac{1}{m_e}\Big) \,.
\end{split}     
\eeq
The terms in the first line do not depend on $\bm{q}^2$. They lead to a contact term in the position space potential, $\propto \delta^3(\bm r)$, and are absorbed into the $d_2$  Wilson coefficient,  contributing to $d_2^\text{inel.}$ in \eqref{eq:d_2:matching:result}.
In contrast, the third term in eq.~\eqref{eq:fin_amp} is non-analytic in $\bm{q}^2$ and features a branch cut beginning at $\bm{q}^2 = 0$, signaling the presence of long-range contributions associated with two-photon intermediate states. In accordance with eq. (\ref{EM_Power_Counting}), the magnetic polarizability  only contributes to the short-distance part of the potential arising from the interactions with relativistic electrons. To the order we are working, it does not contribute to the long-range potential arising from the interaction with non-relativistic electrons. Using similar methods, we can show that the magnetic polarizability gives rise to long-range potential at higher orders \cite{toappear}.

To isolate this long-range potential, one can use dispersion formalism \cite{Feinberg:1989ps}. The Fourier transform of a non-analytic term like $\sqrt{-\bm{q}^{2} - i0}$ is performed using a subtracted dispersion relation. Defining $t = q^2$, the potential in coordinate space is given by \cite{Ghosh:2024ctv}
\begin{align}\label{eq:dispersion}
    V(r) = \frac{1}{8\pi^2 r} \int_{t_0}^\infty dt\, e^{-\sqrt{t} r} \, \text{Disc}[i\mathcal{M}(t)],
\end{align}
where $t_0$ denotes the threshold of the branch cut, typically $t_0 = 0$ for massless intermediate states, and $\text{Disc}[i\mathcal{M}(t)]$ represents the discontinuity across the cut. The discontinuity and the imaginary part are related via
\beq
\text{Disc}[i\mathcal{M}(t)] = -2 \Im[\mathcal{M}(t)], 
\eeq
where
\beq
\Im[(x - i0)^k] = \theta(-x)(-x)^k\sin\left(-k\pi\right).
\eeq

Evaluating the integral in eq.~\eqref{eq:dispersion} using this identity leads to a $1/r^4$ long-range potential,\footnote{The $q$ independent terms in \eqref{eq:fin_amp} lead to  $V_\text{contact}=-({\alpha\, m_e}/{4\pi}) \left[\left(5\,\alpha_E-\beta_M\right)\log\left({\Lambda}\big/{m_e}\right)+4\,\alpha_E\right]\delta^3(\bm r) $, where $\Lambda$ denotes an ultraviolet cutoff associated with the excitation scale of the nucleus. The delta-function term appearing at $\mathcal{O}(m_e)$ contributes to the energy shifts of $s$-states. A similar result was obtained in refs.~\cite{Khriplovich:1997fi, Eides:2000xc, Hill:2012rh}, involving a two-photon exchange without momentum transfer. }
\beq\label{eq:pol_contribution}
    V_\text{pol.}(r)= 
    -\frac{\alpha}{8\pi}\frac{\alpha_E}{r^4}+ \mathcal{O}\left(\frac{1}{m_e}\right) \,.
\eeq
We obtain the same expression for the potential as in refs.~\cite{Ericson:1972nhh, Eides:2000xc}; however, the form of the long range potential was obtained in~\cite{Ericson:1972nhh} by arguing that, if the potential for $\bm{E}$ goes as $1/r^2$, then it should go as $1/r^4$ for $\bm{E}^{\,2}$. Here, we have  derived analytically the potential from the amplitude of a two-photon exchange due to polarizability.

\subsection{One-loop matching: other contributions}
\label{sec:one-loop:other}

In addition to the one-loop contribution from the two-photon polarizability vertex, which we calculated in sec.~\ref{sec:polarizability}, there are also two-photon one-loop graphs that are generated by two insertions of single-photon coupling of the nucleus (see Figs. \ref{fig:Min_Min}, \ref{fig:r2}, \ref{fig:r22}  in app.~\ref{Appendix:Box}, which also contains further details). 
None of these contributions result in a new long-range potential:

\begin{itemize}
\item[(i)] Two minimal $A^0$ couplings ($\propto Z^2$): 
In the non-relativistic/potential region these graphs reproduce the iteration of the Coulomb interaction and are accounted for by solving the Schr\"odinger equation with $V_C(r)=-Z\alpha/r$; they do not match onto a new short-distance potential, see, e.g., refs.~\cite{Pineda:1998kn,Hill:2011wy}.

\item[(ii)] One minimal and one $\langle r^2\rangle$ coupling ($\propto Z\,\langle r^2\rangle$): 
The $\langle r^2\rangle$ vertex supplies a factor of $\bm q^{\,2}$ that cancels one of the photon propagators. The result is analytic in $\bm q^{\,2}$ and renormalizes local dimension-8 contact operators in ${\cal L}_{S\psi_e}$.

\item[(iii)] Two $\langle r^2\rangle$ couplings ($\propto \langle r^2\rangle^2$): 
Both photon propagators are canceled, reducing the graphs to bubble-type integrals with static denominators.
In the on-shell static limit they vanish in dimensional regularization, see app.~\ref{Appendix:Box}.
\end{itemize}

At one-loop order, there are also contributions involving insertions of the unique dimension-six\footnote{Contributions from dimension-eight four-field operators are power suppressed.} four-field operator ${\cal L}_{d_2}$.  These also do not lead to long-range potentials:

\begin{itemize}
\item[(iv)] Insertion of one four-field operator and one Coulomb photon: The resulting diagram involves only a single massless Coulomb propagator, that carries the external momentum. 
Without a pair of propagators sharing the momentum flow, there is no pinch singularity as $\bm q^{\,2}\!\to\!0$, and therefore the amplitude is  analytic in $\bm q^{\,2}$, 
renormalizing the short-distance coefficients in eqs.~(\ref{L_S}) and (\ref{contact_interactions}).
 
 \item[(v)]  Double insertions of four-field contact interactions: Their effect is to renormalizes $d_2$. They  do not introduce a new instantaneous potential or any non-analytic $\bm q$-dependence.  
\end{itemize}

In short, at one loop the only long distance potential is generated by the two-photon polarizability contribution, and is given in eq.~\eqref{eq:pol_contribution}.

\section{Higher order corrections to energy levels} 
\label{sec:energy_levels}

\subsection{Recapitulation of relevant QM potentials}
Before we calculate the King NL from the EFT approach, we collect the full list of QM potentials that we have derived. Apart from the Coulomb potential itself, we have two QM potentials arising from EFT one-photon exchange diagrams (see sec~\ref{sec:Tree}), 
\begin{equation}
V_{\langle r^2\rangle} (\bm r)=\frac{4\pi Z\alpha}{6}\,\langle r^2\rangle\,\delta^3(\bm r)\,,
\end{equation}
\begin{equation}
V_{\langle r^4\rangle} (\bm r) = \frac{4\pi Z\alpha}{120}\,\langle r^4\rangle\,\nabla^2\big[\delta^3(\bm r)\big]\,.
\end{equation}
From contact interactions (see sec.~\ref{sec:contact}), we have three spin-independent QM potentials, from dimension-6 contact interaction 
\begin{equation}
V_{d_2}(\bm r)=- \frac{d_2 m_e}{\Lambda_N^3}\,\delta^3(\bm r) \,,
\end{equation}
from dimension-8 contact interactions,
\begin{align}
V_{d_3}(\bm r)&=\frac{d_3}{\Lambda_N^4}\nabla^2\left[\delta^3(\bm r)\right] , 
\\
V_{d_6}(\bm r)&=-\frac{d_6}{m_e \Lambda_N^3}\big(\overleftarrow{\nabla}^2 \delta^3(\bm r)+\delta^3(\bm r) \nabla^2\big).
\end{align}
We also have the $1/r^4$ potential from 1-loop contribution due to electric polarizability,
\beq
    V_\text{pol.}(r)= 
    -\frac{\alpha}{8\pi}\frac{\alpha_E}{r^4}.
\eeq

We can estimate the relative size of the various short-distance contributions using
first-order QM perturbation theory. Since
$\int d^3r\,|\phi|^2=1$, the electronic bound-state wave function scales as $|\phi| \sim a^{-3/2}$, where $a$ is the atomic length scale. For hydrogen-like systems, $a=(Z m_e\alpha)^{-1}$, 
and we define $\epsilon \equiv r_N/a \sim 10^{-3} (10^{-2})$ as the ratio of nuclear to atomic scales, with the numerical values given for Ca (Yb) \cite{Angeli:2013epw,ParticleDataGroup:2024cfk}.
The finite-size corrections to the energy levels from $\langle r^2\rangle$ are then parametrically, 
\beq
\Delta E_{\langle r^2\rangle}
\sim |\phi(0)|^2\,Z \alpha\langle r^2\rangle
\sim \frac{Z \alpha\langle r^2\rangle}{a^3}
\sim \frac{Z \alpha}{a}\,\epsilon^2,
\eeq
consistent with the proton-radius effects being suppressed by $\epsilon^2$ relative to the
atomic binding energy $  Z^2 \alpha^2 m_e=Z\alpha/a$.
Proceeding similarly, the parametric size of energy shift $\Delta E_i$ induced by each potential is 
\begin{subequations}\label{scaling}
\begin{align}
\langle r^2\rangle &:~~ (Z \alpha/a)\,\epsilon^2, &
\qquad
\langle r^4\rangle &:~~ (Z\alpha/a)\,\epsilon^4,
\\
\label{eq:d2:d3}
d_2 m_e/\Lambda_N^3 &:~~ (Z\alpha/a)\,(Z\alpha)^2\,\epsilon^3, &
\qquad
d_3/\Lambda_N^4 &:~~ (Z\alpha/a)\,(Z\alpha)^3\,\epsilon^4,
\\
\label{eq:d6:polariz}
d_6/m_e \Lambda_N^3 &:~~ (Z\alpha/a)\,(Z\alpha)^2\,\epsilon^3,
&\qquad \text{Polarizability} &:~~ (Z\alpha/a)\,Z^{2/3} \alpha\,\epsilon^3, &
\end{align}
\end{subequations}

In above scaling estimates, we assumed that: (i) all non-perturbative parameters scale as powers of the nuclear size assumed in \eqref{contact_interactions}, which we already know is only approximate --- for $d_2$ there are corrections that can be an order of magnitude larger, see \eqref{eq:d_2:matching:result}; (ii) same as $d_2$, also $d_3$ and $d_6$ start at order $Z^2\alpha^2$, though this is not yet known; and (iii) for electric polarizability we assumed the heavy nuclei scaling in hydrodynamic model,  $\alpha_E\propto Z^{5/3}\alpha$, see eq.~\eqref{eq: elect_pol_1}.
 
Next, we use these potentials to find the shift to the energy levels and from them, determine the King NL. 

\subsection{First order in QM perturbation theory}
\label{sec:1st:order:QM}
The energy level shifts in the first-order QM perturbation theory are given by the expectation values of the various QM potentials, 
\beq
\Delta E_i=\langle V_i \rangle_\phi.
\eeq
We split the evaluation of the expectation values for $s$-levels  and higher levels. 

{\bf $\bm s$-levels:}  Two of the potentials, $V_{\langle r^2\rangle}$ and $V_{d_2}$, include $\delta^3(\bm{r})$ without any derivatives. The $\langle V_i\rangle_\phi$ expectation values for these involve the integral
\begin{equation}
\int d^3 \bm{r} \,\phi^*(\bm{r})\delta^3(\bm{r})\phi(\bm{r})
=\left|\phi(0)\right|^2.
\end{equation}
Since the wave function of a bound state in a central potential with angular momentum $\ell$ scales as $r^\ell$ for small $r$, these potentials yield non-zero contributions only for $s$-levels. Explicitly,  
\begin{align}\label{eq:delta_pot}
\Delta E^{\,s-\mbox{\scriptsize{level}}}_{\langle r^2\rangle}&=\frac{4\pi Z\alpha}{6}\,\langle r^2\rangle\,\left|\phi(0)\right|^2, 
\\
\Delta E^{\,s-\mbox{\scriptsize{level}}}_{d_2}&= -\frac{d_2 m_e}{\Lambda_N^3}\,\left|\phi(0)\right|^2. 
\end{align}

The potentials that depend on $\langle r^4\rangle$ and \(d_3\) include $\nabla^2[\delta^3(\bm{r})]$, and thus $\langle V_{\langle r^4\rangle}\rangle_\phi,\,\langle V_{d_2}\rangle_\phi$ require the integral 
\beq
\begin{split}
\label{eq:r4_d3_potential}
 \int d^3 \bm{r} \,\phi^*(\bm{r}) &\left[\nabla^2\delta^3(\bm{r})\right]  \phi(\bm{r})
=\int d^3 \bm{r} \,\nabla^2\left|\phi(\bm{r})\right|^2\delta^3(\bm{r})
=\nabla^2\left|\phi(0)\right|^2
\\
&=\left[\nabla^2\phi(0)^*\right]\phi(0)+2\,\bm\nabla\phi(0)^*\cdot \bm\nabla\phi(0)+\phi^*(0)\left[\nabla^2\phi(0)\right].
\end{split}
\eeq
Assuming a central potential, this vanishes for $\ell>1$ states. 

For $s$-levels, where $\phi$  is real, there is a divergence in the $r\to 0$ limit as noted previously in the literature, see, e.g., \cite{Lepage:1997cs, Stryker:2015ika}. We can isolate the divergence
\beq
\nabla^2\left|\phi(\bm{r})\right|^2\bigg|_{r \to 0}
=6\phi(0)\phi^{\prime\prime}(0)+6\left[\phi^{\prime}(0)\right]^2+\dfrac{4\phi(0)\phi^{\prime}(0)}{r},
\eeq
where the prime denotes a derivative with respect to $r$, and we used that the $s$-level wave function $\phi$ is real. 
This divergence is UV in nature since it arises from the $r\to 0$ limit. We expect it to cancel against an opposite term in the EFT calculation. This can be checked by using a full single-photon nuclear potential, $\delta V(\bm{r})$ and analyzing the integral $\int d^3 \bm{r} \,\phi^*(\bm{r})\delta V(\bm{r})\phi(\bm{r})$ in momentum space. We find that large-momentum IR divergence cancels against the small-momentum UV divergence (see also app.~\ref{Appendix:UV_IR_divergencies} and ref.~\cite{toappear}). For our purposes we only note these divergent terms, but will not consider them in the King NL analysis. The energy level shift due to \( V_{\langle r^4\rangle},\, V_{d_2}\) with a UV cutoff \(r_{\rm UV}\) is thus
\begin{align}\label{eq:DeltaE:r4:s}
\Delta E^{\,s-\mbox{\scriptsize{level}}}_{\langle r^4\rangle}=&\frac{4\pi Z\alpha}{20}\,\langle r^4\rangle\,\left\{\phi(0)\phi^{\prime\prime}(0)+\left[\phi^{\prime}(0)\right]^2+\frac{2\phi(0)\phi^\prime(0)}{3\,r_{\rm UV}}\right\},
\\
\label{eq:DeltaE:r4:d3:s}
\Delta E^{\,s-\mbox{\scriptsize{level}}}_{d_3}=& \frac{6\,d_3}{\Lambda_N^4}\,\left\{\phi(0)\phi^{\prime\prime}(0)+\left[\phi^{\prime}(0)\right]^2+\frac{2\phi(0)\phi^\prime(0)}{3\,r_{\rm UV}}\right\}.
\end{align}

The expectation value for the $V_{d_6}$ potential involves the integral 
\beq
\begin{split}
I_{d_6}=\int d^3\mathbf r\,
\Big\{\phi^*(\mathbf r)\,\nabla^2\phi(\mathbf r)+ \big[\nabla^2\phi^*(\mathbf r)\big]\phi(\mathbf r)\Big\}\delta^3(\mathbf r)&=\phi^*(0)\,\nabla^2\phi(0)+
\big[\nabla^2\phi(0)\big]^*\,\phi(0) 
\end{split}
\eeq
This is non-zero only for the \(s-\)level (for \(p-\)level and higher \(\ell-\)states, the wave function \(\phi\) vanishes at the origin) and has the form
\begin{equation}
I_{d_6}=2\,\phi(0)\,\nabla^2\phi(0)=6\phi(0)\phi^{\prime\prime}(0)+4\left[\phi^{\prime}(0)\right]^2+4\dfrac{\phi(0)\phi^{\prime}(0)}{r}\bigg|_{r \to 0},
\end{equation}
where in the first equality we used the fact that for $s-$levels the wave functions is real.
The $s-$level energy shift given a UV cutoff \(r_{\rm UV}\) for the divergent piece is
\begin{eqnarray}
\Delta E_{{d_6}}^{\,s-\text{level}}&=&-\frac{4\,d_6}{m_e\Lambda_N^3}\left\{\frac{3\,\phi(0)\phi^{\prime\prime}(0)}{2}+\left[\phi^{\prime}(0)\right]^2+ \frac{\phi(0)\phi^\prime(0)}{r_{\rm UV}}\right\}.
\end{eqnarray}

Finally, the energy level shift from the $1/r^4$ potential due to the electric polarizability is given by
\begin{equation}
\Delta E_\text{pol.} = -\frac{\alpha\,\alpha_E}{8\pi}\int d^3 \bm{r} \,\phi^*(\bm{r})\frac1{r^4}\phi(\bm{r}).
\end{equation}
For the $s$-levels, this results in $1/r_\text{UV}$ and $\ln r_\text{UV}$ divergences when using $r_\text{UV}$ as the UV cut-off on the integral \cite{Eides:2000xc}. For a hydrogen-like atomic system we find as a function of $n$ is then
\begin{equation}\label{eq:pol_l0}
\begin{split}
    \Delta E^{\ell=0}_\text{pol.} &= -\frac{\alpha\,\alpha_E}{8\pi} \int_{r_{\rm UV}}^\infty dr\, \frac{[R_{n,0}(r)]^2}{r^2}
    \\
    &= -\frac{\alpha\,\alpha_E}{8\pi}\frac{8(Z\alpha m_e)^4}{n^3}\left\{\frac{1}{2Z\alpha m_e r_{\rm UV}}+\ln \left(\frac{2Z\alpha m_e r_{\rm UV}}{n}\right)+\gamma_E+C_n\right\},
    \end{split}
\end{equation}
where $C_n$ are rational numbers that can be calculated from the coefficients of the Laguerre polynomial in $R_{n,0}$. For example, $C_1=-1$, $C_2=-3/8$, and $C_3=1/54$. 
Though we do not show this explicitly, we expect all such divergences to be systematically canceled by the properly renormalized Wilson coefficients of higher dimension operators \cite{toappear}, and we thus set the divergent parts to zero. 

{\bf Higher $\bm \ell$-levels:} For $V_{\langle r^4\rangle}$ and $V_{d_3}$, for $p$-level, from eq.~\eqref{eq:r4_d3_potential} only the $2\,\bm\nabla\phi(0)^*\cdot \bm\nabla\phi(0)$ results in a non-zero expectation value. Using the standard definition $\phi_{n1m}(\bm{r})=R_{n1}(r)\,Y_{1m}(\theta,\varphi)$, we find the contribution to energy shift due to \(\langle r^4 \rangle,\,d_3\) to be
\begin{align}
\label{eq:DeltaE:r4:p}
\Delta E^{\,p-\mbox{\scriptsize{level}}}_{\langle r^4\rangle}&=\frac{4\pi Z\alpha}{80\pi}\,\langle r^4\rangle\,\left[R_{n1}^\prime(0)\right]^2,\\
\Delta E^{\,p-\mbox{\scriptsize{level}}}_{d_3}&=\frac{3\,d_3}{2\pi\Lambda_N^4}\,\left[R_{n1}^\prime(0)\right]^2.
\end{align} 
For $\ell>1$ the contributions to $\Delta E^{\,p-\mbox{\scriptsize{level}}}, \Delta E^{\,p-\mbox{\scriptsize{level}}}$ vanish.

For the non-\(s\) level, the contribution to the energy shift from \(1/r^4\) potential in electric polarizability is finite and given by,
\begin{equation}\label{eq:pol_l_level}
    \Delta E^{\ell\neq0}_\text{pol.} = -\frac{\alpha\,\alpha_E}{8\pi} \int_0^\infty dr\, \frac{[R_{n,\ell}(r)]^2}{r^2}.
\end{equation}
For a hydrogen-like atomic system, the energy shift as a function of \(n,\ell\) is then \cite{Eides:2000xc}
\begin{equation}
     \Delta E^{\ell\neq0}_\text{pol.} =-\dfrac{\alpha\,\alpha_E(Z\,\alpha\,\mr)^4}{16\pi\,n^5}\dfrac{3\,n^2-\ell\,(\ell+1)}{\ell(\ell+1)(\ell+3/2)(\ell^2-1/4)}.
\end{equation}

\subsection{The $\langle r^2 \rangle^2$ contribution from second-order in QM perturbation theory}
\label{sec:r2:2}

At second-order in QM perturbation theory we only need to evaluate the contributions from $V_{\langle r^2\rangle}$, to match the current level of experimental precision. These contributions are nonzero only for $s-$level, where we have for the $n$-th energy level, see, e.g., ref.~\cite{Friar:1978wv}, 
\beq
\begin{split}
\label{rsquaredsquared} 
\Delta E^{(2)}_n&=\int d^3\bm{r}\,d^3{\bm r^\prime}\phi_n(\bm r)\,V_{\langle r^2\rangle}(\bm r) \,\ReducedG(\bm r,\bm{r^\prime},E_n)\,\,V_{\langle r^2\rangle}(\bm{r^\prime}) \phi_n(\bm{r^\prime})
\\
&=\left(\frac{4\pi Z\alpha}{6}\right)^2\langle r^2\rangle^2\, \phi_{n,0}(0)\,\ReducedG(0,0,E_n)\, \phi_{n,0}(0),
\end{split}
\eeq
where $\ReducedG$ is the ``reduced" Green's function of the system.\footnote{In ref.~\cite{Friar:1978wv} this was denoted as $G^\prime$.} It is given by 
\beq
\ReducedG=G-\frac{1}{(E-E_n)} \phi_{n,0}(\bm r) \phi_{n,0}(\bm{r^\prime}),
\eeq
where $G$ is the Green's function 
QM system that satisfies (equating $m_e$ with the reduced mass as in the rest of the paper)
\begin{equation}
\label{Greens}
\left[-\nabla^2/(2m_e)+V(r)-E\right]G(\bm r,\bm{r^\prime},E)=\delta(\bm{r}-\bm{r}^{\,\prime})\,.
\end{equation}

Since the bound state energy scales as $Z\alpha/a$, the reduced Green's functions scales as $\ReducedG\sim 1/\left(Z\alpha\,a^2\right)$. Equation (\ref{rsquaredsquared}) then implies that the energy level correction $\Delta E^{(2)}_n$ scales as $(Z\alpha/a)\,\epsilon^4$. This is the same size as the energy shift due to $\langle r^4\rangle$, eqs.~\eqref{eq:DeltaE:r4:s} and \eqref{eq:DeltaE:r4:p}.    
Note, however, that unlike the energy level corrections that arose at first order in QM perturbation theory that we derived in sec.~\ref{sec:1st:order:QM}, the  energy shift $\Delta E^{(2)}_n$ does not depend just on the energy level that is being perturbed, but rather on the entire spectrum of the Hamiltonian. That is, calculating the  $\langle r^2 \rangle^2$ contribution to the energy levels requires the knowledge of the full Green's function. This can be obtained by expanding in $\alpha$ and $\epsilon$ the potential $V(r)$, and similarly the energy levels and the Green's function \eqref{Greens}, obtaining a system of differential equations each containing only terms of the same size in terms of $\alpha$ and $\epsilon$ scaling. 
The reduced Green's function is thus given by
\beq
\label{eq:G:split}
\tilde G=G^{(0)}+\tilde G_Z+\cdots,
\eeq
where 
\beq
G^{(0)}=-2m_e/4\pi|\bm r-\bm{r^\prime}|,
\eeq
solves the ${\mathcal O}(\alpha^0, \epsilon^0)$ differential equation, $-\nabla^2/(2m_e)G^{(0)}=\delta(\bm{r}-\bm{r}^{\,\prime})$, and $\tilde G_Z$ is the reduced Green's function solution of the corresponding ${\mathcal O}(\alpha, \epsilon^0)$ differential equation, in which the potential is the Coulomb one, $V(r)=V_Z(r)$. The ellipses in \eqref{eq:G:split} denote contributions that are further suppressed by powers of $\alpha$ and/or $\epsilon$, and that can be ignored at our precision. For $\tilde G_Z$, we use the results from ref.~\cite{Friar:1978wv}, for which  in our calculation, eq.  (\ref{rsquaredsquared}), we only need the $s$-wave part,
\begin{equation}
4\pi \tilde G_Z=-\frac{2\mr}{r_>}\Big\{1-Z\alpha m_r r -Z\alpha m_r r^\prime-2Z\alpha\mr r_>\big[\log(2Z\alpha\mr r_>)+2\gamma_E-\frac32+\psi(n)-\frac1n-\ln n\big]\Big\}, 
\end{equation}
with $r_>=\text{max}(r,r^\prime)$, $\psi(n)$ the digamma function, $\gamma_E$ the Euler-Mascheroni constant, and $m_r\approx m_e$ in our case. In fact, for King NL we only need the $n$-dependent part of the above expression, while the $n$-independent part can be absorbed into the field shift. That is, writing $\tilde G_Z$ as a sum of $n-$dependent and $n-$independent terms,  $\ReducedG_Z=\ReducedG_Z^{n\text{-indep.}}+\ReducedG_Z^{n\text{-dep.}}$, we have for the $n$-dependent part, 

\begin{equation}
\ReducedG_Z^{n\text{-dep.}}(0,0,E_n)=\frac{m_e^2(Z\alpha)}{\pi}\left[\psi(n)-\frac1n-\ln n\right].
\end{equation}

In the calculation of $\Delta E^{(2)}_n$ the universal part of the Green's function, $G^{(0)}$, as well as the $n-$independent part of the Coulomb reduced Green's function, $\ReducedG_Z^{n\text{-indep.}}$, give a contribution that is proportional to $|\phi_{n,0}(0)|^2$. As we will show below, in sec. \ref{NL_EFT}, such terms can be absorbed into effective $\langle r^2 \rangle$ contribution to the energy shifts, including the divergence\footnote{Note that the $G^{(0)}$ contribution in eq.~\eqref{rsquaredsquared} is also UV divergent. In the formalism of ref.~\cite{Friar:1978wv} this term gave rise to the Friar radius $\langle r^3\rangle_{(2)}$, while in our formalism the Friar radius contribution arises from $d_2$.} in (\ref{rsquaredsquared}) from $\ln r_>$ term in $\ReducedG_Z^{n\text{-indep.}}$. The $\ReducedG_Z^{n\text{-dep.}}$ term, on the other hand, gives a different dependence on $n$ and thus contributes to King NL, as we discuss in more detail in the next section.

\section{Predictions for nonlinearities}
\label{sec:pheno}

\subsection{Calculating nonlinearities from EFT}
\label{NL_EFT}
Let us now apply the above results to the calculation of King NL using EFT. The energy shifts due to finite nuclear effects have the general form 
\begin{equation}
\label{eq:EFT_energy_shift}
\Delta E_{n,\ell}=F_{n,\ell}\, \langle r^2 \rangle_\text{eff}+G_{n,\ell}^{(2)}\,\langle r^2 \rangle^2+G_{n,\ell}^{(4)}\,\langle r^4\rangle_{\mbox{\scriptsize eff}}+P_{n,\ell}\, \alpha_E+D_{n,\ell} \, \d6tilde+\cdots\,,
\end{equation}
where we explicitly denoted the dependence on the EFT Wilson coefficients that encode nuclear effects, $\langle r^2 \rangle_\text{eff}$, $\langle r^2 \rangle^2$, $\langle r^4\rangle_{\mbox{\scriptsize eff}}$, $\alpha_E$, and $\d6tilde\equiv d_6/(m_e\Lambda_N^3)$, while the prefactors $F(n,\ell)$, $G_2(n,\ell)$, $G_4(n,\ell)$, $P(n,\ell)$, and $D(n,\ell)$ depend only on electron wave function, and thus on principal quantum number $n$ and angular momentum $\ell$. The expressions for these prefactors are given in equations (\ref{eq:IS_contributionsF})-(\ref{eq:IS_contributionsD}) and (\ref{Pnell_H}). The ellipsis denotes higher order effects and the mass shift discussed below.

The nuclear size effects that result in the same electron wave function dependence were lumped together in \eqref{eq:EFT_energy_shift}  into effective coefficients, $\langle r^2 \rangle_\text{eff}$ and $\langle r^4\rangle_\text{eff}$. At leading order in $\epsilon$, i.e., at ${\mathcal O}(1/\Lambda_N^2)$, $\langle r^2 \rangle_\text{eff}$ coincides with $\langle r^2 \rangle$, while including corrections up to ${\mathcal O}(1/\Lambda_N^4)$ one has for hydrogen-like atoms (see also the discussion below),
 \beq
 \begin{split}
 \label{eq:r2:eff}
 \langle r^2 \rangle_{\text{eff}}=& \langle r^2 \rangle-\frac{6m_e}{4\pi Z\alpha}\frac{d_2}{\Lambda_N^3}
+\langle r^2 \rangle^2\left(\frac{4\pi Z\alpha}6\right)\ReducedG_Z^{n\text{-indep.}}+\\
&+\left(m_eZ_\text{eff} \alpha\right)^2\left[\frac3{10}\langle r^4 \rangle+\frac{9}{\pi Z\alpha} \frac{d_3}{\Lambda_N^4} - \frac{6}{\pi Z\alpha} \frac{d_6}{m_e\Lambda_N^3}\right].
\end{split}
\eeq
The coefficient $\langle r^4 \rangle _\text{eff}$, on the other hand, receives corrections from higher dimensional contact  interaction operators already at leading order in $\epsilon$, 
\beq
\langle r^4 \rangle_\text{eff}= \langle r^4 \rangle+\frac{30}{\pi Z\alpha}\frac{d_3}{\Lambda_N^4}-\frac{30}{\pi Z\alpha}\frac{d_6}{m_e\Lambda_N^3}.
\eeq
The definitions of the EFT operators corresponding to the Wilson coefficients $\langle r^2\rangle$, $\langle r^4\rangle$, $d_2$, $d_3$, and $d_6$ can be found in eqs.~\eqref{L_S} and \eqref{contact_interactions}.

Next, let us estimate the size of King NL.
Consider an atomic transition $i=(n,\ell)\to (n^\prime,\ell')$ with frequency $\nu_i^A$ for isotope $A$, and the same transition, but now for a different isotope, $A'$, so that the transition frequency\footnote{To convert energy ($E$) to frequency ($\nu$) we need to divide by $h$, the Planck constant. Since we are using $\hbar=1$ units, we have $E=\omega=2\pi\nu$. } is $\nu_i^{A'}$. 
The difference, $\omega_i^{AA^\prime}\equiv2\pi\nu_i^{AA^\prime}\equiv 2\pi(\nu_i^A-\nu_i^{A^\prime})$, is then given by, cf. eq.~\eqref{eq:EFT_energy_shift},
\begin{equation}
\label{eq:nuiAA'}
\begin{split}
\omega_i^{AA^\prime}=&K_i \,\mu_{AA'}+F_i\left(\langle r_A^2 \rangle_{\mbox{\scriptsize eff}}-\langle r_{A^\prime}^2 \rangle_{\mbox{\scriptsize eff}}\right)+G^{(2)}_i\left(\langle r_A^2 \rangle^2-\langle r_{A^\prime}^2 \rangle^2\right)
\\
&+G^{(4)}_i\left(\langle r_A^4 \rangle_\text{eff}-\langle r_{A^\prime}^4 \rangle_\text{eff}\right)+P_i\big(\alpha_{E,A}-\alpha_{E,A'}\big)+
\cdots,
\end{split}
\end{equation}
where $K_i$ is the electronic coefficient multiplying the mass shift, $\mu_{AA'}=1/M_A-1/M_{A'}$, while $F_i=F_{n,\ell}-F_{n^\prime,\ell'}, G^{(2/4)}_i=G^{(2/4)}_{n,\ell}-G^{(2/4)}_{n^\prime,\ell'}$, and 
$P_i=P_{n,\ell}-P_{n',\ell'}$.
The ellipses denote the contribution proportional to $\d6tilde$ and higher order terms. We leave the full evaluation of $\d6tilde$ for future work, since it requires a matching calculation for dimension 8 operators. Given that we have not yet performed the matching for these higher dimension operators, it is even possible that the $d_6$ contribution is zero to the order we are working.

The largest contributions to $\nu_i^{AA'}$ in \eqref{eq:nuiAA':delta}  are from $\mu_{AA'}$ and $\langle r_A^2 \rangle_\text{eff}$ terms. If these are the only two nuclear size/mass effects, one can write down a linear relation between two atomic transitions, $i=1,2$, 
\beq
\label{eq:KingNL:general}
\frac{\omega_2^{AA^\prime}}{\varepsilon_{AA'}}=\frac{F_2}{F_1}\frac{\omega_1^{AA^\prime}}{\varepsilon_{AA'}}+\frac{1}{M_A}\Big(K_2-\frac{F_2}{F_1}K_1\Big)+\Delta_\text{NL}^{AA'},
\eeq
where we introduced a small parameter 
\beq
\varepsilon_{AA'}\equiv M_A \mu_{AA'}\simeq \frac{A'-A}{A}, 
\eeq
while $\Delta_\text{NL}^{AA'}$ contains non-linear corrections. That is, for $\Delta_\text{NL}^{AA'}=0$ there is a so called King linearity relation between the reduced frequency differences, $\nu_{1,2}^{AA^\prime}/\varepsilon_{AA'}$, with the same proportionally factor, $F_2/F_1,$ and a constant off-set, $\big[K_2-K_1(F_2/F_1)\big]/M_A$, for any pair of isotopes $A, A'$. 

The non-linear correction to the King linearity relation is given by
\beq
\begin{split}
\label{eq:Delta:NL:AA:full}
\varepsilon_{AA'} \,\Delta_\text{NL}^{AA'}=&\Big(G_2^{(2)}-\frac{F_2}{F_1}G_1^{(2)}\Big)\left(\langle r_A^2 \rangle^2-\langle r_{A^\prime}^2 \rangle^2\right)
+\Big(G_2^{(4)}-\frac{F_2}{F_1}G_1^{(4)}\Big)\left(\langle r_A^4 \rangle_\text{eff}-\langle r_{A^\prime}^4 \rangle_\text{eff}\right)
\\
&+\Big(P_2-\frac{F_2}{F_1}P_1\Big)\big(\alpha_{E,A}-\alpha_{E,A'}\big)+\cdots.
\end{split}
\eeq
Note that this correction vanishes either in the limit where nuclear size effects are isotope independent, or in the limit of all the electronic factors proportional to the field shift ones, i.e., $G_2^{(2)}/G_1^{(2)}={F_2}/{F_1}$,... That is, King NL requires effects that differ \emph{both} in the electronic and the nuclear factors. Assuming naive scalings with $A$ for the nuclear size effects: $\Lambda_N\propto A^{-1/3}$, $\langle r^2\rangle\propto A^{2/3}$, $\langle r^4\rangle\propto A^{4/3}$, and using  the large mass scaling for polarizability in the hydrodynamic model,  $\alpha_E\propto A^{5/3}$, the King NL correction is then given by 
\beq
\begin{split}
\label{eq:nuiAA':delta} 
\Delta_\text{NL}^{AA'}\simeq&\frac{\left(\varepsilon_{AA'}\right)^0}{3}\biggr[-4 \Big(G_2^{(2)}-\frac{F_2}{F_1}G_1^{(2)}\Big) \langle r_A^2 \rangle^2-4\Big(G_2^{(4)}-\frac{F_2}{F_1}G_1^{(4)}\Big)\langle r_A^4 \rangle_\text{eff}-5 \Big(P_2-\frac{F_2}{F_1}P_1\Big) \alpha_{E,A}\biggr]\\&+\frac{\varepsilon_{AA'}}{9}\biggr[-2 \Big(G_2^{(2)}-\frac{F_2}{F_1}G_1^{(2)}\Big) \langle r_A^2 \rangle^2-2\Big(G_2^{(4)}-\frac{F_2}{F_1}G_1^{(4)}\Big)\langle r_A^4 \rangle_\text{eff}
\\&\qquad\quad-5 \Big(P_2-\frac{F_2}{F_1}P_1\Big) \alpha_{E,A}\biggr]+\cdots.
\end{split}
\eeq
Note that $\Delta_\text{NL}^{AA'}$ we can drop the $\left(\varepsilon_{AA'}\right)^0$ terms, since they only correct the constant off-set in the King NL relation, while ellipses denote terms of higher order in $\varepsilon_{AA'}$. 
The expression in \eqref{eq:nuiAA':delta} is only approximate, and for precise prediction one should use \eqref{eq:Delta:NL:AA:full} with nuclear size coefficients extracted from other data. Eq.~\eqref{eq:nuiAA':delta} does highlight the fact that King NL first arises at ${\mathcal O}(\epsilon^3)$ and ${\mathcal O}(\epsilon^4)$ from three distinct types of nuclear structure effects. Parametrically, the different contributions in \eqref{eq:nuiAA':delta}  are (see also the more detailed calculation for hydrogen-like atoms, given below)
\begin{align}
\langle r^2\rangle^2&: \quad 
(Z\alpha/a)\epsilon^4,
\\
\langle r^4\rangle_\text{eff} &:  \quad 
(Z\alpha/a)\epsilon^4,
\\
\alpha_{E} &: \quad (Z\alpha/a)\,Z^{2/3}\alpha\, \epsilon^3,
\end{align}
where as illustration we used the same scalings for Wilson coefficients as in eqs.~\eqref{eq:d2:d3} and \eqref{eq:d6:polariz}. The not yet evaluated $d_6$ contribution, which we do not display in \eqref{eq:nuiAA':delta},  is parametrically of size $(Z\alpha/a)(Z\alpha)^2 \epsilon^3$
and thus naively expected to be small. 

To gauge the sizes of different contributions to King NL, let us calculate the expressions for electronic prefactors in the expression for the energy levels, \eqref{eq:EFT_energy_shift}, for a hydrogen-like system. For this, we only need the expressions for wave functions $\phi_{n,\ell}(r)$ that are solutions of the Schr\"odinger equation for the Coulomb potential $V(r)= -Z_\text{eff}\,\alpha/r$.\footnote{Corrections to $\phi_{n,\ell}(r)$ from nuclear size effects amount to subleading corrections. Note also that we use $Z_\text{eff}$ instead of $Z$, to be able to trace which $Z$ factors arise from wave-function and which from the definitions of the operators. In the numerical analysis we can set $Z=Z_\text{eff}$.} The finiteness of the Schr\"odinger equation for $r\rightarrow0$ implies,
 \begin{equation}
\phi_{n,0}^{\prime}(0)=-m_e Z_\text{eff} \,\alpha \, \phi_{n,0}(0).
\end{equation}
This relation holds also for \(V(r) = -Z_\text{eff}\alpha/r+V_\text{regular}(r)\), where $V_\text{regular}(0)=0$. We find, 
\begin{align}\label{eq:IS_contributionsF}
F_{n,\ell}&=\frac{4\pi Z\alpha}6\left|\phi_{n,\ell}(0)\right|^2\, \delta_{\ell 0},
\\
G^{(2)}_{n,\ell}&=\left(\frac{4\pi Z\alpha}6\right)^2\ReducedG_{Z_\text{eff}}^{n\text{-dep.}}(0,0,E_n)\,\left|\phi_{n,\ell}(0)\right|^2\,\delta_{\ell 0},
\\
G^{(4)}_{n,\ell}&= \begin{cases}
    \dfrac{4\pi Z\alpha}{20}\phi_{n,0}(0)\phi_{n,0}^{\prime\prime}(0), & \ell =0, \\[2mm]
    \dfrac{4\pi Z\alpha}{80\pi}\left[R_{n1}^\prime(0)\right]^2, & \ell =1,\\[2mm]
0, & \ell > 1,
\end{cases}\\
\label{eq:IS_contributionsD}
D_{n,\ell}&=\left\{-6\phi(0)\phi_{n,\ell}^{\prime\prime}(0)-4\left[\phi_{n,\ell}^{\prime}(0)\right]^2\right\}\, \delta_{\ell 0},.
\end{align}

For completeness we show also the expression for $D_{n,\ell}$. For hydrogen-like atom the wave-function at $r=0$ satisfy 
\begin{align}
\left|\phi_{n,0}(0)\right|^2&=
\frac{(Z_\text{eff} \alpha m_e)^3}{n^3\pi}\,,  
\\
\label{eq:phi_second_derivative}
\phi_{n,0}(0)\phi_{n,0}^{\prime\prime}(0)&= 
\frac{(Z_\text{eff} \alpha m_e)^5}{3n^5\pi}+\frac{2(Z_\text{eff} \alpha m_e)^5}{3n^3\pi}\,,
\end{align}
thus giving
\begin{align}\label{eq:IS_contributions_hydrogen}
F_{n,\ell}&=\frac{2(Z \alpha)(Z_\text{eff} \alpha m_e)^3}{3n^3}\,\delta_{\ell 0},
\\
G_{n,\ell}^{(2)}&=\dfrac{4m_e^5(Z \alpha)^2(Z_\text{eff} \alpha)^4}{9n^3}\left[\psi(n)-\frac1n-\ln n\right]\,\delta_{\ell 0},
\\
G^{(4)}_{n,\ell}&= \begin{cases}
    \dfrac{(Z \alpha)(Z_\text{eff} \alpha m_e)^5}{15n^5}, & \ell =0, \\[2mm]
    \dfrac{(Z \alpha)(Z_\text{eff} \alpha m_e)^5}{45n^3}-\dfrac{(Z \alpha)(Z_\text{eff} \alpha m_e)^5}{45n^5}, & \ell =1,\\[2mm]
0, & \ell > 1,
\end{cases}
\\
D_{n,\ell}&=- \dfrac{2(Z_\text{eff} \alpha m_e)^5}{n^5\pi}\,\delta_{\ell 0},
\end{align}
where in $G^{(4)}_{n,0}$ and $D_{n,0}$ we assume that the $1/n^3$ terms can be absorbed by a redefinition of the field shift. The electronic prefactor for the polarizability term is 
\beq\label{Pnell_H}
    P_{n,\ell} = 
    \begin{cases}
        -\dfrac{\alpha\,\alpha_E}{8\pi}\dfrac{8(Z_\text{eff}\alpha m_e)^4}{n^3}\left\{\ln \left(\dfrac{2Z_\text{eff}\alpha m_e}{n}\right)+\gamma_E+C_n\right\}, & \ell =0, \\[5mm]
        -\dfrac{\alpha\,\alpha_E(Z_\text{eff}\alpha m_e)^4}{16\pi\,n^5}\dfrac{3\,n^2-\ell\,(\ell+1)}{\ell(\ell+1)(\ell+3/2)(\ell^2-1/4)}, & \ell\neq 0.
    \end{cases}
\eeq
In the expression above for \(\ell=0\) from eq.~\eqref{eq:pol_l0}, we only show the finite terms that contribute to the energy shift due to polarizability, after the divergent pieces have been canceled. 
The electronic coefficient multiplying the mass shift \(K_i\) is
\begin{equation}
K_{n,\ell} = 2\pi R_\infty c \frac{Z^2_\text{eff}\,m_e}{n^2}= \frac{Z^2_\text{eff}\,m_e^2(c\,\alpha)^2}{\,2\, n^2}\,,   
\end{equation}
where $R_\infty$ is the Rydberg constant \cite{ParticleDataGroup:2024cfk}. 

We can use the above results for numerical estimates. 
In table~\ref{tab: IS_contributions}, we first list in the 2nd column the differences of transition frequencies, $\nu_i^{AA'}$, taking $A'=A+2$, for three different transitions, $\alpha$: $3S \to 2P$,  
$\beta$: $4D \rightarrow 3P$, and  
$\gamma$: $3D \rightarrow 2S$.
In the numerical analysis we set $Z_\text{eff}=Z$ and use a simple-minded model for the Wilson coefficients given by a uniform-sphere model for the nucleus, and ignoring higher order terms in  $\langle r^2\rangle_\text{eff}$ and $\langle r^4\rangle_\text{eff}$, so that
\beq
\langle r^2\rangle_\text{eff}=\langle r^{2} \rangle = \frac{3}{5}R^2,\qquad  \langle r^{2} \rangle^{2} = \Big(\frac{3}{5}R^2\Big)^2, \qquad \langle r^{4} \rangle_\text{eff} = \langle r^{4} \rangle = \frac{3}{7}R^4,
\eeq
where $R = r_0\,A^{1/3}$. In table~\ref{tab: IS_contributions}, 2nd column, we then show the leading dependence on $A$ and $Z$, both for $\nu_i^{AA'}$, and for different contributions to it, as given in eq.~\eqref{eq:nuiAA'}. In 3rd to 5th column, we then give the numerical examples for the case of $\text{Ca}^{19+}$ isotopes, taking as examples $(A,A')=\{(40,42),(42,44), (44,46)\}$, while always $Z=20$. For $\alpha_E$ we use eq.~\eqref{eq:elect_pol_2}. For polarizability contribution to $n$S-levels we only use the $n$ dependent terms in  eq.~\eqref{Pnell_H} since the $n$-independent terms can be absorbed into the field shift. The numerical values for \(C_n\) in eq.~\eqref{Pnell_H} for the transitions we are interested in are \(C_2, C_3 = -3/8, 1/54\). 

\begin{table}[t]
\begin{threeparttable}
\centering
\begin{tabular}{@{}llccccc@{}}
\toprule\toprule
 & H-like; $A'=A+2$ & $^{40,42}\text{Ca}^{19+}$ & $^{42,44}\text{Ca}^{19+}$ & $^{44,46}\text{Ca}^{19+}$ \\
\midrule 
$\nu_{\alpha=3S \to 2P}$  
 & $-4.6\times 10^{2} \,Z^2$ THz & $-1.8 \times 10^{5}$ THz & $-1.8 \times 10^{5}$ THz & $-1.8 \times 10^{5}$ THz   \\ \\[-3mm]
$\nu_{\beta=4D \to 3P}$ &  $-1.6\times 10^{2}\,Z^2$ THz & $-6.4 \times 10^{4}$ THz & $-6.4 \times 10^{4}$ THz & $-6.4 \times 10^{4}$ THz  \\ \\[-3mm]
$\nu_{\gamma=3D \to 2S}$ & $-4.6\times 10^{2} \,Z^2$ THz & $-1.8 \times 10^{5}$ THz & $-1.8 \times 10^{5}$ THz & $-1.8 \times 10^{5}$ THz    \\
\midrule \midrule
$K_{\alpha}\,\mu_{AA^\prime}/h$ & $5.0 \times 10^{2}\,Z^2/A^2$ GHz & $1.3\times 10^{2}$ GHz & $1.1\times 10^{2}$ GHz & $1.0\times 10^{2}$ GHz \\ \\[-3mm]
$K_{\beta}\,\mu_{AA^\prime}/h$ & $1.8\times 10^{2}\,Z^2/A^2$ GHz & $4.4\times10^1$ GHz & $4.0\times10^1$ GHz & $3.6\times10^1$ GHz\\ \\[-3mm]
$K_{\gamma}\,\mu_{AA^\prime}/h$ & $5.0\times 10^{2} \,Z^2/A^2$ GHz & $1.3\times 10^{2}$ GHz & $1.1\times 10^{2} $ GHz & $1.0\times 10^{2}$ GHz  \\
\midrule 
$F_{\alpha}\,\langle r^2\rangle_{AA^\prime}/h$ & $-6.1\times10^{-5}\,Z^4/A^{1/3}$ GHz & $-2.9$ GHz & $-2.8$ GHz & $-2.8$ GHz  \\ \\[-3mm]
$F_{\beta}\,\langle r^2\rangle_{AA^\prime}/h$ & 0 &0 &0 & 0 \\ \\[-3mm]
$F_{\gamma}\,\langle r^2\rangle_{AA^\prime}/h$ & $2.1\times10^{-4}\,Z^4/A^{1/3}$ Hz & $9.7$ GHz & $9.5$ GHz & $9.4$ GHz    \\
\midrule 
$G^{(2)}_{\alpha}\,\langle r^2\rangle^{2}_{AA^\prime}/h$ & $1.2\times10^{-8}\,Z^6\,A^{1/3}$ kHz & $2.6$ kHz & $2.6$ kHz & $2.7$ kHz \\ \\[-3mm]
$G^{(2)}_{\beta}\,\langle r^2\rangle^{2}_{AA^\prime}/h$ & 0 & 0& 0 &0\\ \\[-3mm]
$G^{(2)}_{\gamma}\,\langle r^2\rangle^{2}_{AA^\prime}/h$ & $-6.0\times10^{-8}\,Z^6\,A^{1/3}$ kHz & $-1.3\times10^{1}$ kHz & $-1.3\times10^{1}$ kHz & $-1.4\times10^{1}$ kHz  \\
\midrule 
$G^{(4)}_{\alpha}\,\langle r^4\rangle_{AA^\prime}/h$ & $3.0\times10^{-9}\,Z^6\,A^{1/3}$ kHz & $6.6\times10^{-1}$ kHz & $6.8\times10^{-1}$ kHz & $6.9\times10^{-1}$ kHz \\ \\[-3mm]
$G^{(4)}_{\beta}\,\langle r^4\rangle_{AA^\prime}/h$ & $1.2\times10^{-9}\,Z^6\,A^{1/3}$ kHz  & $2.7\times10^{-1}$ kHz & $2.8\times10^{-2}$ kHz & $2.8\times10^{-1}$ kHz\\ \\[-3mm]
$G^{(4)}_{\gamma}\,\langle r^4\rangle_{AA^\prime}/h$ & $3.5\times10^{-9}\,Z^6\,A^{1/3}$ kHz & $7.6\times10^{-1}$ kHz & $7.8\times10^{-1}$ kHz & $7.9\times10^{-1}$ kHz \\
\midrule 
$P_{\alpha}\,\alpha_E^{AA^\prime}/h$ & $-6.4\times10^{-8}\,Z^4\,A^{2/3}$ MHz & $-1.2\times10^{-1}$ MHz & $-1.2\times10^{-1}$ MHz & $-1.3\times10^{-1}$ MHz\\ \\[-3mm]
$P_{\beta}\,\alpha_E^{AA^\prime}/h$ & $-2.4\times10^{-9}\,Z^4\,A^{2/3}$ MHz & $-4.5\times10^{-3}$ MHz & $-4.6\times10^{-3}$ MHz & $-4.7\times10^{-3}$ MHz \\ \\[-3mm]
$P_{\gamma}\,\alpha_E^{AA^\prime}/h$ & $1.9\times10^{-7}\,Z^4\,A^{2/3}$ MHz & $3.5\times10^{-1}$ MHz & $3.7\times10^{-1}$ MHz & $3.8\times10^{-1}$ MHz\\
\bottomrule \bottomrule
\end{tabular}
\caption{Numerical example of isotope shift frequency differences for several  transitions,
$\alpha$: $3S \rightarrow 2P$,  
$\beta$: $4D \rightarrow 3P$, and  
$\gamma$: $3D \rightarrow 2S$,  in hydrogen-like systems, taking $A'=A+2$ (2nd column) and for three isotope combinations for the case of Ca$^{19+}$. Different rows show either the total frequency shift (with \(h\) being the Planck's constant), or different contributions to it as in \eqref{eq:nuiAA'}, where we shortened $\langle r^2\rangle_{AA^\prime}=\langle r^2\rangle_{A}-\langle r^2\rangle_{A'}$, etc.}
\label{tab: IS_contributions}
\end{threeparttable}
\end{table}

\subsection{Comparison to Friar's predictions for  $s-$levels in hydrogen-like atoms}
\label{sec:friar}
It is illuminating to compare our EFT based results with the purely QM calculation for hydrogen-like atoms \cite{Friar:1978wv, Borisoglebsky:1979nb}. In ref.~\cite{Friar:1978wv} the nuclear finite-size effects were calculated  for $s$-levels in hydrogen-like atoms using QM perturbation theory, and assuming that the effect of nucleus can be described fully by a classical charge distribution $\rho$ centered at $r=0$. 
The results for the shifts to the energy levels were expressed in terms of moments of a general (unknown) charge distribution, e.g., $\langle r^n\rangle_\rho=\int d^3 r \,\rho\, r^n /Ze$, and are (in the $M\to \infty$ limit)
\begin{equation}
\label{eq:DeltaEn:Friar}
\Delta E_{n,0}^\text{QM}=\frac{2 \pi}{3} |\phi_{n,0}(0)|^2 Z \alpha \left[\;
  \langle r^2 \rangle_\rho\,\, { -\frac{ Z \alpha m_e  }{2}\,\langle r^3\rangle_{\rho(2)}}
		    +(Z \alpha m_e)^2 F_{\mbox{\scriptsize NR}} +(Z \alpha )^2F_{\mbox{\scriptsize REL}}\right],
\end{equation}
with 
\begin{align}
\begin{split}
F_{\mbox{\scriptsize NR}} =&{ \dfrac{2}{3} \Big(\psi(n)-\frac{1}{n}-\ln n\Big)\langle r^2 \rangle_\rho^2+ \dfrac{1}{10n^2}\langle r^4\rangle_\rho}+
\dfrac{2}{3}\Big(2 \gamma-\dfrac{4}{3}\Big)  \langle r^2 \rangle_\rho^2        
             \\
&+\dfrac{2}{3}\langle r^2\rangle_\rho \langle r^2 \ln(2Z\alpha m_e r)\rangle_\rho+ \langle r^3 \rangle_\rho \langle r \rangle_\rho+ \dfrac{1}{9}\langle r^5 \rangle_\rho \langle r^{-1} \rangle_\rho
             + I_2^{\mbox{\scriptsize NR}} + I_3^{\mbox{\scriptsize NR}},
  \end{split}           
             \\
\begin{split}             
F_{\mbox{\scriptsize REL}} =&  -\langle r^2 \rangle_\rho \Big(\psi(n)+2\gamma+ \dfrac{9}{4n^2}-\frac1n-\dfrac{13}{4}
               + \ln(2Z\alpha m_e r) -\ln n\ \Big)
 \\
               &-\dfrac{1}{3}\langle r^3 \rangle_\rho\langle r^{-1}\rangle_\rho
               +  I_2^{\mbox{\scriptsize REL}}+ I_3^{\mbox{\scriptsize REL}},
\end{split}               
\end{align}
where $I_{2,3}^{\mbox{\scriptsize NR,REL}}$ do not depend on $n$, see ref.~\cite{Friar:1978wv} for their explicit forms.
Since the above calculation assumes a classical charge distribution it cannot capture all the nuclear finite-size effects. For example, the term  $|\phi_n(0)|^2 Z \alpha Z \alpha m_e\sim {\cal O}\big(Z\alpha(m_e Z \alpha)^4\big)$ in \eqref{eq:DeltaEn:Friar}, receives corrections due to nuclear excitations that are not included in the above expression. To understand such differences between the two results in more detail, let us 
rearrange the expression \eqref{eq:DeltaEn:Friar} into a form 
that makes the comparison with the EFT result in \eqref{eq:EFT_energy_shift} straightforward, 
\begin{equation}
\label{eq:energy_shift}
\Delta E_{n,0}^\text{QM}=F_{n,0}^\text{QM}\langle r^2 \rangle_{\rho,\text{eff}}+G_{n,0}^{(2)\text{QM}}\langle r^2 \rangle_\rho^2+G_{n,0}^{(4)\text{QM}}\langle r^4\rangle_\rho+{\cal O}\big((Z \alpha)^7\big).
\end{equation}
Using $|\phi_{n,0}(0)|^2=(Z \alpha m_e)^3/(\pi n^3)$, the electronic pre-factors are given by
\begin{align}\label{eq:Friar_HG2G4}
F_{n,0}^\text{QM}&=\frac{2(Z \alpha)(Z \alpha m_e)^3}{3n^3}\left\{1+(Z \alpha)^2\Big(-\psi(n)-\dfrac{9}{4n^2}+\frac1n+\ln n\Big)\right\},
\\
G_{n,0}^{(2)\text{QM}}&=\frac{4(Z \alpha)(Z \alpha m_e)^5}{9n^3}\Big(\psi(n)-\frac1n-\ln n\Big),
\\
G_{n,0}^{(4)\text{QM}}&=\frac{(Z \alpha)(Z \alpha m_e)^5}{15n^5},
\end{align}
while the effective charge radius square is 
 \beq
 \begin{split}
 \label{eq:r2:eff:rho}
 \langle r^2 \rangle_{\rho, \text{eff}}=& \langle r^2 \rangle_\rho\,\, { -\frac{ Z \alpha m_e  }{2}\,\langle r^3\rangle_{\rho(2)}}+(Z \alpha)^2 m_e^2\bigg[\frac23\langle r^2 \rangle_\rho^2\left(2 \gamma-\dfrac{4}{3}\right)+\langle r^3 \rangle_\rho \langle r \rangle_\rho
 \\
 &+\langle r^2\rangle_\rho \langle r^2 \ln(2Z\alpha m_e r)\rangle_\rho + \dfrac{1}{9}\langle r^5 \rangle_\rho \langle r^{-1} \rangle_\rho
             + I_2^{\mbox{\scriptsize NR}} + I_3^{\mbox{\scriptsize NR}}\bigg]
             \\
&+(Z \alpha)^2\bigg[\left(-2\gamma+\dfrac{13}{4}\right) \langle r^2 \rangle_\rho-\langle r^2\rangle_\rho \ln(2Z\alpha m_e r)
\\
&-\frac{1}{3}\langle r^3 \rangle_\rho\langle r^{-1}\rangle_\rho
               +  I_2^{\mbox{\scriptsize REL}}+ I_3^{\mbox{\scriptsize REL}}\bigg].  
\end{split}                
\eeq

Comparing eq.~\eqref{eq:energy_shift} with the EFT result in eq. \eqref{eq:EFT_energy_shift}, we notice several differences. First of all, while the expressions for the electronic prefactors $G_{n,\ell}^{(2)}$ and $G_{n,\ell}^{(4)}$ agree when restricted to $s$-levels, the expressions for $F_{n,\ell}$ differ, due to relativistic corrections that were not included in our EFT calculations. While these corrections could be included also in the EFT calculation,  the difference is not important in practice for the measurements of King NL, since the constant offset in King linearity relation, eq.~\eqref{eq:KingNL:general}, is fitted from data.

The second difference is that, even when $\langle r^2\rangle$ and $\langle r^4\rangle$ EFT Wilson coefficients are interpreted as the corresponding moments of a classical nuclear charge distribution, $\langle r^2\rangle_\rho$ and $\langle r^4\rangle_\rho$, the expressions for $\langle r^2 \rangle_\text{eff}$ in eq.~\eqref{eq:r2:eff} and $\langle r^2 \rangle_{\rho,\text{eff}}$ in eq.~\eqref{eq:r2:eff:rho} differ at orders $\left(Z\alpha\right)$ and $\left(Z\alpha\right)^2$. That is, at ${\mathcal O}(Z\alpha)$ the Friar radius contribution in $\langle r^2 \rangle_\text{eff}$ is only a part of the $d_2$ Wilson coefficient entering $\langle r^2 \rangle_\text{eff}$, and contains both the inelastic contributions from nuclear excitations as well as one loop contributions from matching at scale $M$, see eq.~\eqref{eq:d_2:matching:result}. At ${\mathcal O}\big((Z\alpha)^2)$ the expression for $\langle r^2 \rangle_{\rho,\text{eff}}$ includes contributions that would correspond to a two-loop matching for $d_2$ Wilson coefficients in the EFT,  which we did not include in this paper, while the EFT result also includes the contribution from $d_3$ and higher dimension operators. While further work is required to fully understand these differences, it is also clear that the $d_3$ contribution is not relevant for King NL since the field shift (dependence on $\langle r^2\rangle_\text{eff}$) gets factored out explicitly when constructing the King linearity relation, eq. \eqref{eq:KingNL:general}.

The third difference is that $\langle r^4 \rangle_\text{eff}$ differs from $\langle r^4 \rangle_\rho$ by the contributions from the $d_3$ operators.  
The difference is irrelevant for King NL that is insensitive to redefinitions of  $\langle r^4 \rangle$. The fourth difference is that polarizability was not considered in \cite{Friar:1978wv}. 
Finally, there is the contribution from the $d_6$ operator, which was also not considered in \cite{Friar:1978wv}. If one limits the discussion to only transitions between $s-$levels, this contribution can be included in $\langle r^4 \rangle$, while in general it constitutes a new contribution to King NL.

\subsection{Comparison with approximate treatments of $\langle r^2\rangle^2$ term}
\label{sec:other:r22}

In sec. \ref{sec:r2:2}, we showed that the evaluation of $\langle r^2\rangle^2$ contribution to the energy shift requires the knowledge of the full QM spectrum. The calculation of the electronic prefactor $G_i^{(2)}$ multiplying the quadratic field shift in \eqref{eq:nuiAA'} is expected to remain rather involved also when considering multi-electron systems. Formally, it corresponds to calculating
\begin{equation}
\label{eq:G2i:correct}
    G^{(2)}_{i} = \frac{1}{2}\frac{\partial^2 \omega_i^{AA'}}{(\partial \langle r_A^2\rangle)^2}.
\end{equation}
Part of the quadratic dependence of $\Delta E_{n,\ell}$ on $\langle r^2\rangle$ is due to the variation of the wave function at the origin as $\langle r^2\rangle $ changes in the linear field shift contribution to the energy levels. Following refs.~\cite{Counts:2020aws,Hur:2022} which for $s-$levels would then give from eq.~\eqref{eq:IS_contributions_hydrogen} the following approximation\footnote{The proportionality factor of ${\pi Z\,\alpha}/{3}$ in \eqref{eq:G2i:short-cut} is different from \cite{Counts:2020aws} since we are using $\hbar=1$ units and define $\rho_{i}$ as the 3-D electronic density, not the radial electronic density.} 
  
\begin{equation}
\label{eq:G2i:short-cut}
    G^{(2)\text{app}}_{i} = \frac{\pi Z\,\alpha}{3} \frac{\partial \rho_{i}(0;\langle r^2 \rangle)}{\partial \langle r^2 \rangle}.
\end{equation}
Here, $\rho_{i}(0;\langle r^2 \rangle)$ is the difference in electronic density at the origin for $i=(n,\ell)\to (n',\ell')$ transition.
Since this is much easier to calculate, refs.~\cite{Counts:2020aws,Hur:2022}, for instance, used it as an estimator of the full $G^{(2)}_{i}$ term in \eqref{eq:G2i:correct}. In fact, to make the calculation 
numerically tractable,  ref.~\cite{Hur:2022} made a further approximation, calculating the derivative in \eqref{eq:G2i:short-cut} as
\begin{equation}
\label{eq:s18}
    \frac{\partial \rho_{i}(0;\langle r^2 \rangle)}{\partial \langle r^2 \rangle} = C_{i}\frac{\partial \rho^P_{i}}{\partial r^2}\Big|_{r^2=\langle r^2\rangle} \,,
\end{equation}
where $\rho_i^P$ is the charge density difference for a point nucleus and $C_i$ a constant, conjectured to be the same as the following ratio of charge density differences
\beq
\label{eq:CiP}
C_{i}^P= \frac{\rho_{i}(0;\langle r^2 \rangle)}{ \rho^P_{i}(r^2 = \langle r^2 \rangle)}.
\eeq

The need for such approximate treatments illustrates, on the one hand, how difficult it is to obtain the sizes of King NL effects for multi-electron systems. While 
both the ratio in \eqref{eq:CiP} and the derivative on r.h.s. of \eqref{eq:s18} are much easier to calculate, on the other hand, it is unclear how well these approximate the full results. This is difficult to answer at present for multielectron systems, however, using our framework we can test how well the approximations work for hydrogen-like atoms. In appendix \ref{sec:app:r22}  we give further details and also analytical comparisons, while here let us give a representative numerical example, taking $n = 2$ and $n' = 3$ with $Z = 1$, $r_p = 0.842$ fm, $m_e = 0.511$ MeV. This gives $G^{(2)\text{app}}_{i}/G^{(2)}_{i}=1.2$ since $G^{(2)\text{app}}_{i}$ includes also terms that can be absorbed into $F_i$. Note, however, that this comparison is highly dependent on how the other moments of classical nuclear charge density are treated, see appendix \ref{sec:app:r22}.
For $C_i^P$ approximating $C_i$, the two are similar in magnitude, though of opposite signs. These results highlight the need for further study on calculating $G^{(2)}_{i}$.

\section{Summary and outlook}\label{Summary}

In this manuscript we developed an effective field theory (EFT) description of King Non-Linearity (King NL) that cleanly separates nuclear and atomic physics. The SM was matched onto scalar NRQED in the $M\!\to\!\infty$ limit, with the nuclear structure encoded in a small set of Wilson coefficients ($\langle r^2\rangle,\langle r^4\rangle,\alpha_E,\beta_M,\ldots$) multiplying higher dimension operators. This EFT description was then matched onto quantum-mechanical (pNRQED) potentials for the bound electron.
 
Using this framework we showed that King NL first arises from four distinct sources at ${\mathcal O}(\epsilon^3)$ and ${\mathcal O}(\epsilon^4)$ in the ratio $\epsilon=r_N/a$, where $a$ is a typical atomic size, and  $r_N$ a typical nuclear size. Two sources can be directly related to a dimension-7 operator and a dimension-8 operator of the non-relativistic EFT, namely, the electric polarizability, $\alpha_E$, and the effective ``quartic"-charge radius, $\langle r^4\rangle_\text{eff}$, respectively. The third source arises at the second order in quantum mechanical perturbation theory, and is proportional to $\langle r^2\rangle^2$, while the fourth source is due to a dimension-8  operator with the Wilson coefficient $d_6$, whose size is at present only conjectured at the parametric level. The $\alpha_E$ and $\langle r^4\rangle_\text{eff}$ contributions require the knowledge of only a few leading electronic levels, while the $\langle r^2\rangle^2$ term requires knowledge of all the levels of the system. Parametrically, polarizability contribution is the dominant one, and enters in King NL at ${\mathcal O}((Z\alpha/a)\,Z^{2/3} \alpha\,\epsilon^3)$, assuming a hydrodynamic model for the size of polarizability, while $\langle r^2\rangle^2$ and  $\langle r^4\rangle_\text{eff}$ contributions are nominally of  $\mathcal O((Z\alpha/a) \epsilon^4)$.

In terms of effective potentials at tree level, we find that the $\langle r^2\rangle$ and $\langle r^4\rangle$ operators result in short-range potentials $V_{\langle r^2\rangle}\propto \delta^3(\bm r)$ and $V_{\langle r^4\rangle}\propto \nabla^2\delta^3(\bm r)$. Finite-size two-photon effects are captured by contact operators: the unique dimension-6 term includes the Friar  radius contribution, while the relevant dimension-8 term acting on the electron produces a local $\nabla^2\delta^3(\bm r)$ structure (finite for $\ell\!\ge\!1$, renormalizing $s$-waves). One-loop graphs with two single-photon couplings are in the static on-shell limit  analytic in $\bm q^{\,2}$ and do not generate a long-range potential. The only  long-range effect at this order is due to nuclear electric polarizability, which at one loop gives rise to a potential with $1/r^4$ behavior as well as a local contact term that renormalizes higher dimension operators. The commonly discussed $\langle r^2 \rangle^2$ contribution arises solely from second-order QM perturbation with a double insertion of $V_{\langle r^2\rangle}$ and the reduced Coulomb Green's function at the origin; this reproduces a quadratic field shift and requires no extra nuclear input beyond $\langle r^2\rangle$. The utility of the EFT approach that we employ is most apparent in clear power counting in $r_N/a$ and $\alpha$  of different $\langle r^2\rangle$, $\langle r^4\rangle$,  polarizability and contact term contributions, allowing to identify the dominant contributions to King NL.

In this work we limited the discussion to scalar nuclei and hydrogen-like systems. There are thus several important effects that need to be incorporated into our theoretical framework in order to be able to match with the current experimental data on King NL. 
First of all, our current treatment needs to be extended to include multi electron systems.  Multi-electron effects are important when calculating transition amplitudes for higher $\ell$ states, and can potentially lead to sizable King NL even in absence of nuclear polarizability. For instance, for calcium it was proposed that the leading effect  could arise from  the quadratic mass shift~\cite{Wilzewski:2024wap}.
 Second, incorporation of higher order relativistic effects will allow for proper treatment of hyperfine transitions and spin-orbit effects. Finally, one would want to move away from the static limit for the nucleus and include effects that are suppressed by powers of $1/M^n$.

\acknowledgments
We thank Roni Harnik and Robert Szafron for useful discussions.
G.L. acknowledges the support of the Samsung Science \& Technology Foundation under Project Number SSTF-BA1601-07, 
a Korea University Grant, and the support of the U.S. National Science Foundation through grant PHY1719877.
G.L. is grateful to the University of Toronto for partial support during completion of this work.
G.~Perez is funded by ISF, Minerva, NSF-BSF and the European Union (ERC, DM-Dawn, 101199868).   
B.A. and J.Z. acknowledge support in part by the DOE grants
DE-SC1019775, DE-SC0026301, and the NSF grants OAC-2103889, OAC-2411215, and OAC-2417682. This work was performed in part at the Aspen Center for Physics, with support for B.A.\ by a grant from the Simons Foundation (1161654, Troyer).

\appendix 
\section{Matching calculation of NRQED Wilson coefficients}
\label{sec:app:matching}

\subsection{One-photon matching of Wilson coefficients} \label{One_photon_Wilson}
\begin{figure}[t]
    \centering  \includegraphics[width=0.25\linewidth]{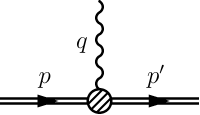}
    \caption{\label{fig:One_Photon_Matching} One-photon interaction with a scalar vertex represented as shaded circle.}
\end{figure}
The one-photon interaction of a scalar is parametrized by a form factor $F(q^2)$, defined via 
\begin{equation}\label{F_defined}
\Gamma^{(3)}=-i e\left(p+p^\prime\right)^\mu F(q^2),
\end{equation} 
where $\Gamma^{(3)}$ is the irreducible three-point function, $q=p^\prime-p$, with $p(p')$ the incoming (outgoing) momentum of the scalar, and $p^2={p^\prime}^2=M^2$, see fig.~\ref{fig:One_Photon_Matching}. For a scalar nucleus, $F(q^2)$ is a non-perturbative object.

We can determine the coefficients of ${\cal L}_S$, eq.~\eqref{L_S:init}, by comparing the full amplitude $\Gamma^{(3)}$ expressed in terms of $F(q^2)$ and the effective theory amplitude,  keeping in mind that the field $S$ in ${\cal L} _S$ is rescaled compared to the standard relativistic field by a factor of $2M$, see, e.g., refs.~\cite{Hill:2011be,Dye:2016uep}. We perform the matching in the rest frame of the initial scalar, i.e., taking $\bm{p}=0$. Comparing the two expressions for Lorentz index $\mu=0$, and using $q^0=-q^2/2M$ as well as $q^2=-\bm{q^2}+{\cal O}\left(1/M^2\right)$ \cite{Dye:2016uep} in eq.~\eqref{F_defined}, we find that the matching condition is 
\begin{equation}
-i e \left[F(0)-F^\prime(0)\bm{q^2}+\frac{1}{2}F^{\prime\prime}(0)\bm{q^4}+\cdots
\right]\stackrel{!}{=}-ieZ+ie\frac{\coneprime}{\Lambda_N^2}\bm{q^2}-ie\frac{\cfourprime}{\Lambda_N^4}\bm{q^4}+\cdots\, ,
\end{equation} 
where the $\stackrel{!}{=}$ sign requires the two sides to be equal. 
This gives 
\begin{equation}
Z=F(0)\,, \quad 
\frac{\coneprime}{\Lambda_N^2}=F^\prime(0) \,, \quad 
\frac{\cfourprime}{\Lambda_N^4}=\dfrac12F^{\prime\prime}(0) \,.
\end{equation} 
Following the standard parameterization for the expansion of the form factor, see, e.g., ref.~\cite{Halzen:1984mc} 
\begin{equation}
    F(\bm q^2)=F(0)\left(1-\frac16\bm{q^2}\langle r^2\rangle+\frac1{120}\bm{q^4}\langle r^4\rangle+\cdots\right),
\end{equation}
we thus have 
\begin{equation}
\langle r^2\rangle=6\dfrac{F^\prime(0)}{F(0)} \,, \quad \langle r^4\rangle=60\dfrac{F^{\prime\prime}(0)}{F(0)} \,.
\end{equation} 
The coefficients of the NRQED Lagrangian \eqref{L_S:init} are thus
\begin{equation}
Z=F(0)\,, \quad 
\frac{\coneprime}{\Lambda_N^2}=F(0)\frac{\langle r^2\rangle}6 \,, \quad 
\frac{\cfourprime}{\Lambda_N^4}=F(0)\frac{\langle r^4\rangle}{120} \,,
\end{equation} 
giving the final expression for ${\cal L}_S$ in eq.~\eqref{L_S} in the main text.

\subsection{Two-photon matching of Wilson coefficients}
\label{Two_photon_Wilson}
\begin{figure}[t]
    \centering  \includegraphics[width=0.25\linewidth]{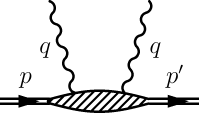}
    \caption{\label{fig:Two_Photon_Matching}Two-photon interaction with a scalar vertex represented as shaded ``blob". }
\end{figure}
Using the result above, we can construct the one photon exchange amplitude between the electron and the nucleus. We find that there is no contribution to the Wilson coefficient $d_2$ in ${\cal L}_{S\psi}$ in eq.~\eqref{contact_interactions} at order $\alpha$. To determine the Wilson coefficient of the dimension-six contact interaction we thus need to consider two-photon exchange forward  amplitude between the electron and the nucleus. 

Let us denote by $W^{\mu\nu}(p,q)$ the two-photon interaction of a scalar nucleus of four-momentum $p$ with a photon of momentum $q$,
\begin{equation}
\label{Wdefined}
W^{\mu\nu}(p,q)=i\int d^4x\, e^{iqx}\langle p|T\left\{J^\mu_{\mbox{\scriptsize e.m.}}(x)J_{\mbox{\scriptsize e.m.}}^\nu(0)\right\}|p \rangle,
\end{equation}
where $J_{\mbox{\scriptsize e.m.}}^\nu=\sum_\psi Q_\psi \bar \psi \gamma^\mu \psi$ is the electromagnetic current in QCD, with $e$ conventionally factored out (the sum is over light quarks, $\psi=u,d,s$, with $Q_\psi$ the charge of quark $\psi$).
Current conservation, 
\begin{equation}
q_\mu W^{\mu\nu}(p,q) = q_\nu W^{\mu\nu}(p,q) = 0 \,,
\end{equation}
implies that $W^{\mu\nu}(p,q)$ can be expressed as 
\begin{equation}\label{Tensordecomposition}
W^{\mu\nu}(p,q) = \left( - g^{\mu\nu} + \dfrac{q^\mu q^\nu}{ q^2} \right) W_1+
\left( p^\mu - \dfrac{p\cdot q \,q^\mu }{q^2} \right) 
\left( p^\nu - \dfrac{p\cdot q \, q^\nu} {q^2} \right) W_2 \,.
\end{equation} 
The two scalar functions $W_{1},W_{2}$ depend on the variables $\nu=2p\cdot q$ and $Q^2=-q^2$.   One can show, see, e.g., \cite{Dye:2018rgg}, that translation invariance implies that $W^{\mu\nu}(p,q)=W^{\nu\mu}(p,-q)$. As a result, $W_1$ and $W_2$ are even functions of $\nu=2p\cdot q$. 
For scalar point particle of charge $+1$, for instance,\footnote{For elementary scalar the definition of $W^{\mu\nu}(p,q) $ in eq.~\eqref{Wdefined} needs to be amended to include contributions from two-photon interactions of the scalar.}
\begin{equation}\label{pointW1W2}
W_1=-2, \qquad W_2=\frac{8Q^2}{Q^4-\nu^2}.
\end{equation}
The expression for $W_2$ is identical to the spin-1/2 case, while $W_1$ is different, see, e.g., \cite{Hill:2011wy}. Unlike the spin-1/2 case, we now have in $W_1$ only the  contribution that does not involve an off-shell point-particle propagator.

In analogy to the spin-1/2 case, we assume that  $W_{1}$ satisfies a subtracted dispersion relation in $\nu$, see, e.g., \cite{Dye:2018rgg},
\begin{equation}
\label{W1disp}
W_1(\nu,Q^2)-W_1(0,Q^2)=\dfrac{2\nu^2}{\pi}\int_{\nu_0}^\infty d\nu^\prime\dfrac{\mbox{Im } W_1(\nu^\prime,Q^2)}{\nu^\prime\left(\nu^{\prime2}-\nu^2\right)},
\end{equation}
and that $W_{2}$ satisfies an unsubtracted dispersion relation
\begin{equation}
\label{W2disp}
W_2(\nu,Q^2)=\dfrac{2}{\pi}\int_{\nu_0}^\infty d\nu^\prime\,\nu^\prime\dfrac{\mbox{Im } W_2(\nu^\prime,Q^2)}{\nu^{\prime2}-\nu^2}.
\end{equation}
As an example, consider the point-particle scalar case. We have $\mbox{Im } W_1=0$ and $\mbox{Im } W_2=4\pi\left[\delta(\nu-Q^2)+\delta(\nu+Q^2)\right]$. Taking  $W_1(0,Q^2)=-2$, we reproduce eq.~\eqref{pointW1W2}. 

From the definition of $W^{\mu\nu}$, eq.~\eqref{Wdefined}, one can obtain the identity \cite{Manohar:2000dt}
\beq
\begin{split}
\label{ImW}
2\,\mbox{Im}\, W^{\mu\nu}=&\sum_X \langle p|J^\mu|X\rangle\langle X |J^\nu|p\rangle (2\pi)^4 \delta^{(4)}\left(q-p_X+p\right)
\\
&+\sum_X \langle p|J^\nu|X\rangle\langle X |J^\mu|p\rangle (2\pi)^4 \delta^{(4)}\left(q+p_X-p\right),
\end{split}
\eeq
where $p_X$ is the four-momentum of the state $X$. The summation includes phase space integrals and sum over spins of the state $X$. This equation allows us to separate the composite scalar and the inelastic contribution in the imaginary part of $W^{\mu\nu}$. The composite scalar contribution is given in terms of a form factor, and the inelastic contribution in terms of inelastic structure function.

Using the dispersion relations we can express the scalar functions as \cite{Hill:2011wy}
\begin{align}\label{decomposition}
W_1(\nu,Q^2)&=W_1(0,Q^2)+W_1^{S,1}(\nu,Q^2)+W_1^{c,1}(\nu,Q^2),
 \\
W_2(\nu,Q^2)&=W_2^{S,0}(\nu,Q^2)+W_2^{c,0}(\nu,Q^2),
\end{align}
where the superscript numbers denote the number of subtractions, the superscript $S$ is the composite scalar contribution, and the superscript $c$ is the continuum, also called inelastic, contribution. 

Let us look at the composite scalar contribution to eq.~\eqref{ImW}. Using (\ref{F_defined}) we have 
\beq
\begin{split}
2 e^2\, \Im W^{\mu\nu} =&\int\frac{d^4p^\prime}{(2\pi)^4}(2\pi)\delta\left[p^{\prime\,2}-M^2\right] \Big\{
\\
&\quad(2\pi)^4 \delta^{(4)}\left(q-p^\prime+p\right)
\left[i e\left(p+p^\prime\right)^\mu F(q^2)\right] \left[-i e\left(p+p^\prime\right)^\nu F(q^2)\right]+
\\
&+ (2\pi)^4 \delta^{(4)}\left(q+p^\prime-p\right)
\left[i e\left(p+p^\prime\right)^\nu F(q^2)\right]\left[-i e\left(p+p^\prime\right)^\mu F(q^2)\right]\Big\}
\\
=&(2\pi)\delta\left(2p\cdot q+q^2\right)e^2\left[F(q^2)\right]^2\left(2p^\mu+q^\mu\right) \left(2p^\nu+q^\nu\right)
\\
&+(2\pi)\delta\left(2p\cdot q-q^2\right)e^2\left[F(q^2)\right]^2\left(2p^\mu-q^\mu\right) \left(2p^\nu-q^\nu\right),
\end{split}
\eeq
which thus gives
\beq
\begin{split}
\Im W^{\mu\nu}
=&4\pi  \left[ F(q^2)\right]^2\left( p^\nu - \dfrac{p\cdot q \, q^\nu} {q^2} \right)\left( p^\mu - \dfrac{p\cdot q \,q^\mu }{q^2} \right)  \left[\delta\left(2p\cdot q+q^2\right)+\delta\left(2p\cdot q-q^2\right)\right]. 
\end{split}
\eeq
We therefore have 
\beq
\Im\,W_1=0, \text{~and~~} \Im\,W_2=4\pi\left[\delta(\nu-Q^2)+\delta(\nu+Q^2)\right]\left[F(q^2)\right]^2.
\eeq
If we take $F(q^2)\equiv 1$ we reproduce the point-particle result. Inserting the expressions for the imaginary parts into the dispersion relations, eqs.~\eqref{W1disp}, \eqref{W2disp}, and assuming that $W_1(0,Q^2)=-2+{\cal O}\left(Q^2\right)$, gives
\begin{equation}\label{eq:W1W2}
W_1=-2+{\cal O}\left(Q^2\right), \quad W_2=\frac{8Q^2}{Q^4-\nu^2}\left[F(-Q^2)\right]^2.
\end{equation}

Next, let us consider the two-photon forward amplitude for  $e^-(k)+S(p)\to e^-(k)+S(p)$ process, in Feynman gauge. In terms of the hadronic tensor we find \cite{Dye:2018rgg}
\begin{equation}\label{full}
i{\cal M}_{\mbox{\scriptsize Full}}=-Z_e^2e^4\int\dfrac{d^4l}{(2\pi)^4}\dfrac{\bar u\gamma_\mu(\kslash-\lslash+m_e)\gamma_\nu u}{(k-l)^2-m_e^2}\left(\dfrac{1}{l^2}\right)^2\frac1{2M}W^{\mu\nu}(p,l), 
\end{equation}
where the $1/2M$ prefactor is due to the non-relativistic normalization of the dimension 3/2 composite scalar field, see, e.g., the appendix of ref.~\cite{Hoang:2005dk}.

To match onto NRQED, we insert eq.~\eqref{decomposition} into eq.~\eqref{full} and set the lepton 3-momentum to zero, i.e., $k=(m_e,\vec 0)$ and work in the rest frame of the nucleus, i.e., $p=(M,\vec 0)$. Furthermore, the electron Dirac spinors should be used with non-relativistic normalization $u(k)^\dagger u(k)=1$, so that we replace $u(k)\to(\chi\,\,0)^T$, where $\chi$ is a two-component spinor. Since $W_1$ and $W_2$ are even functions of $\nu=2p\cdot q=2Mq^0$, this allows to simplify the integration by discarding odd power of $l^0$ in the integral.  After these simplifications we find
\begin{equation}\label{full-v2}
i{\cal M}_{\mbox{\scriptsize Full}}=-Z_e^2e^4\int\dfrac{d^4l}{(2\pi)^4}\dfrac{1}{l^4-(2m_el^0)^2}\left(\dfrac{1}{l^2}\right)^2\frac{m_e}{M}\Big[-W_1\left(l^2+2l_0^2\right)+M^2(l^2-l_0^2)W_2\Big].
\end{equation}
We now perform a Wick rotation $l^0\to iQ u$, where $Q$ is the magnitude of the 4-d Euclidean vector and $u$ the polar angle between $Q$ and the Euclidean time direction. We find
\beq
\begin{split}
\label{fullEuclidian}
{\cal M}_{\mbox{\scriptsize Full}}=&-\frac{m_e}{M}\frac1{4\pi^3}Z_e^2e^4\int_0^\infty dQ \int_{-1}^{1}du\,\sqrt{1-u^2}\frac{Q^3}{\left(Q^2\right)^2}\frac1{Q^2+4m_e^2u^2}\times
\\
&\times\left[\left(1+2u^2\right)W_1(2MiQu,Q^2)-M^2(1-u^2)W_2(2MiQu,Q^2)\right].
\end{split} 
\eeq 
Next, let us specialize to composite scalar contribution. Inserting the expression for $W_2$ from eq.~\eqref{eq:W1W2} and using the fact that $W_1$ is only non-zero at  order $M^0$, we find after the $u$ integration 
\begin{equation}
{\cal M}_{\mbox{\scriptsize Full},S}^{M\to\infty}=16\,m_e\,\alpha^2Z_e^2\, \int_0^\infty \frac{dQ}{{\left(Q^2\right)^2}} \left[F(-Q^2)\right]^2+{\cal O}\left(\frac1{M}\right).
\label{eq:toFri}
\end{equation}
This amplitude has an IR divergence that cancels in the matching onto NRQED. 

Using NRQED Feynman rules we compute the amplitude for photon scattering off a scalar nucleus. In the scalar NRQED propagator we keep the $M$ dependence since $q^0$ is of the same order as $-\bm{q^2}/2M$, see, e.g., \cite{Dye:2018rgg}. For the vertices we only keep the leading interaction in $M$. We find 
\beq
\begin{split}
\label{W2EFT}
   M^2 W_2^{\rm EFT} (\nu, Q^2) &= \dfrac{8Q^2}{Q^4-\nu^2}\left(Z^2\,M^2-\frac14\,Z\,c_D\,Q^2\right)+{\cal O}\left(\frac{Q^2}{M^2}\right)\\
   &=M^2\dfrac{8Q^2}{Q^4-\nu^2}\left(Z^2-\frac13\,Z^2\,\langle r^2 \rangle\,Q^2\right)+{\cal O}\left(\frac{Q^2}{M^2}\right).
\end{split}
\eeq
We can check this result by taking the limits $\nu\to0$ and $Z\to 1$ to reproduce $M^2 W_2^{\rm EFT} (0, Q^2)$ calculated in  \cite{Hill:2012rh}.

The $\nu$ dependence in (\ref{W2EFT}) is identical to that of the full theory. Subtracting the EFT expression, $M^2 W_2^{\rm EFT} (\nu, Q^2)$, from the full theory result for $M^2 W_2(\nu, Q^2)$ given in eq.~\eqref{eq:W1W2}, gives the higher dimension contribution. That is,  equating ${\cal M}_{\mbox{\scriptsize Full},S}^{M\to\infty}-{\cal M}_{\mbox{\scriptsize EFT}}^{M\to\infty}$ to $d_2 m_e/\Lambda_N^3$ gives 
\begin{equation}
\frac{d_2m_e}{\Lambda_N^3}=16\,m_e\,\alpha^2Z_e^2\, \int_0^\infty \frac{dQ}{Q^4} \left\{\left[F(-Q^2)\right]^2-\left[F(0)\right]^2+\dfrac{\left[F(0)\right]^2 \langle r^2 \rangle Q^2}{3}\right\}+{\cal O}\left(\frac1{M}\right).
\end{equation}
Using the relation between $\langle r^2 \rangle$ and $F^\prime(0)$ it is straightforward to check that the integral is IR finite.  It is proportional to the Friar radius (cubed), which is given by 
\beq\label{eq:Friar_radius}
       \langle r^3\rangle_{(2)}=\frac{48}{\pi}\, \frac1{\left[F(0)\right]^2}\int_0^\infty \frac{dQ}{Q^4} \left\{\left[F(-Q^2)\right]^2-\left[F(0)\right]^2+\dfrac{\left[F(0)\right]^2 \langle r^2 \rangle Q^2}{3}\right\}\,,
   \eeq
   so  that, up to inelastic and $1/M$ contributions (see eq. \eqref{eq:d_2:matching:result}),
   
\begin{equation}
\frac{d_2 m_e}{\Lambda_N^3}= \frac{\pi}{3}\,(Z\alpha)^2\,m_e\,\langle r^3\rangle_{(2)}\, .
\end{equation} 
   
The expression in \eqref{eq:Friar_radius} agrees with \cite{Borah:2020gte,Antognini:2022xoo}, if we identify $F(q^2)$ with $G_E(q^2)$. Notice that the analogous expression in ref.~\cite{Eides:2000xc} is missing the $\langle r^2 \rangle Q^2$ term. Given that refs.~\cite{Borah:2020gte,Antognini:2022xoo} do not derive $\langle r^3\rangle_{(2)}$, while the derivation in \cite{Eides:2000xc} is missing a term, the above result appears to be the first complete derivation of the Friar radius using field theory methods.

\section{Matching onto QM potentials} 
\label{Appendix:Box}  
In this appendix, we outline the calculation of the one-loop box diagrams with two exchanged photons.
A detailed study is left for a future work \cite{toappear}.  
\subsection{The $Z^2$ contributions}
\begin{figure}[ht]
    \centering
    \includegraphics[width=0.45\linewidth]{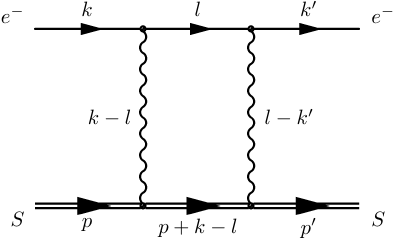}
    \hspace{1cm}
    \includegraphics[width=0.45\linewidth]{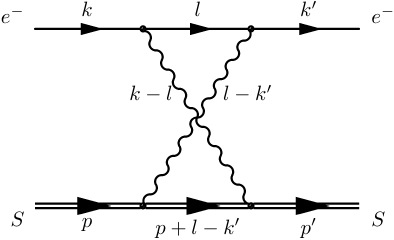}
    \caption{Two-photon exchange with two minimal $A^0$ couplings ($\propto Z^2$). In pNRQED these graphs correspond to the iteration of the Coulomb potential.}
    \label{fig:Min_Min}
\end{figure}
The box and crossed-box diagrams in fig.~\ref{fig:Min_Min} contain two minimal $A^0$ couplings from the $S^\dagger iD_t S$ operator. These reproduce the iteration of the Coulomb potential in the potential region and therefore do not match onto a new short-distance operator. In pNRQED language, integrating out hard/soft modes leaves the instantaneous Coulomb potential $V_C(r)=-Z\alpha/r$. Repeated exchange is generated by solving the Schr\"odinger equation, i.e., by ladder resummation in the potential region, see, e.g., pNRQED and potential-region factorization in refs.~\cite{Pineda:1998kn,Hill:2011wy}.

\subsection{The $Z-\langle r^2\rangle$ contributions}
\begin{figure}[ht]
    \centering
    \includegraphics[width=0.45\linewidth]{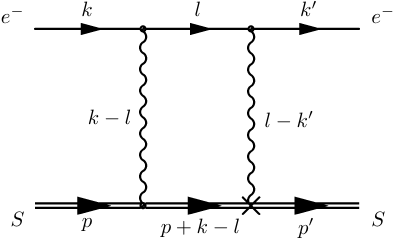}
    \hspace{1cm}
    \includegraphics[width=0.45\linewidth]{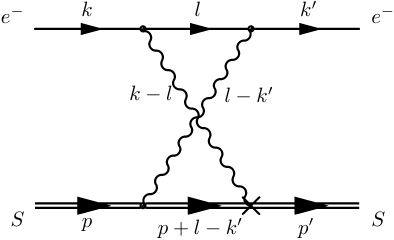}
    \caption{Two-photon exchange with one minimal $A^0$ coupling and one $\langle r^2\rangle$ coupling.}
    \label{fig:r2}
\end{figure}

In the diagrams in fig.~\ref{fig:r2} the $\langle r^2\rangle$ vertex supplies a factor of $\bm q^{\,2}$.
In either topology (left or right panel) this factor cancels one Coulomb propagator, so the box effectively reduces to a triangle with a static propagator.
In the static on-shell limit there is no pinch singularity at $\bm q^{\,2}\to 0$, and the remaining dependence on external kinematics is through $v\!\cdot\!k$ or $v\!\cdot\!k'$, not through $k\!\cdot\!k'$.
Hence the amplitude is analytic in $\bm q^{\,2}$ and generates no nonanalytic (or long-range) piece.
This analytic dependence is expected to be absorbed into local dimension-8 contact operators in ${\cal L}_{S\psi_e}$ (the $d_{3,4,5,6}$ structures), i.e., it renormalizes short-distance coefficients rather than producing a new long-range potential \cite{Pineda:1998kn,Hill:2011wy,Hill:2012rh}.
The same conclusion holds for the crossed graph in the right panel in fig.~\ref{fig:r2}: after the propagator cancellation it depends only on $\bm k^{\prime\,2}$ and likewise renormalizes local operators.

\subsection{The $\langle r^2\rangle^2$ contribution}
\begin{figure}[ht]
    \centering
    \includegraphics[width=0.45\linewidth]{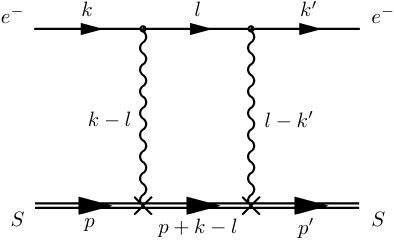}
    \hspace{1cm}
    \includegraphics[width=0.45\linewidth]{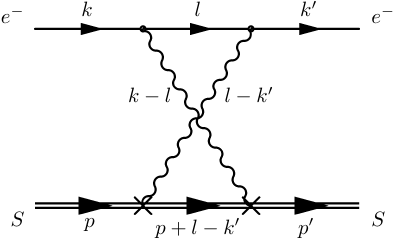}
    \caption{Two-photon exchange with two $\langle r^2\rangle$ couplings.}
    \label{fig:r22}
\end{figure}
The amplitude for the graphs in fig.~\ref{fig:r22} with NRQED vertices
$S^\dagger\, e\,\frac{\langle r^2\rangle}{6}\,[\bm\nabla\!\cdot\!\bm E]\,S$
is
\beq
\begin{split}
i\,\mathcal M_{\langle r^2\rangle^2}
&= \int\!\frac{d^4 l}{(2\pi)^4}\;
\bigg[\bar u(k')\,(-i e \gamma^0)\,
\frac{i}{(\bm l-\bm k')^2}\,
\frac{i(\slashed l+m_e)}{l^2-m_e^2+i0}\,
\frac{i}{(\bm k-\bm l)^2}\,
(-i e \gamma^0)\,u(k)\bigg] 
\\
&\quad\times
\bigg[i e\,|\bm l-\bm k'|^2\; i e\,|\bm k-\bm l|^2\,
\frac{\langle r^2\rangle^2}{36}\,
\bigg(\frac{-1}{v\!\cdot\! l - v\!\cdot\!(p+k)}
+\frac{1}{v\!\cdot\! l - v\!\cdot\!(k'-p)}\bigg)\bigg] 
\\
&=(4\pi\alpha)^2\,\mu^{2\epsilon}\,\frac{\langle r^2\rangle^2}{36}
\int\!\frac{d^d l}{(2\pi)^d}\;
\bigg[\bar u(k')\,\gamma^0\,\frac{\slashed l+m_e}{l^2-m_e^2}\,\gamma^0\,u(k)\bigg] 
\\{}& \quad\times
\bigg(\frac{-1}{v\!\cdot\! l - v\!\cdot\!(p+k)}
+\frac{1}{v\!\cdot\! l - v\!\cdot\!(k'-p)}\bigg).
\label{eq:r22-start}
\end{split}
\eeq
Each $\langle r^2\rangle$ vertex supplies a factor of $\bm q^{\,2}$ that cancels a Coulomb propagator, so the box reduces to a bubble with a single static propagator.
It is convenient to express eq.~\eqref{eq:r22-start} in terms of heavy-light two-point functions:
\beq
\begin{split}
i\,&\mathcal M_{\langle r^2\rangle^2}
=(4\pi\alpha)^2\,\frac{\langle r^2\rangle^2}{36}\,
\bar u(k')u(k)\,m_e\,
\Big[\bar B_0\!\big(m_e,\,v\!\cdot\!k'-v\!\cdot\!p\big)
-\bar B_0\!\big(m_e,\,v\!\cdot\!k+v\!\cdot\!p\big)\Big] 
\\
&\quad+(4\pi\alpha)^2\,\frac{\langle r^2\rangle^2}{36}\,
\bar u(k')\,\gamma^0\gamma^\mu\gamma^0\,u(k)\,
\Big[\bar B_\mu\!\big(m_e,\,v\!\cdot\!k'-v\!\cdot\!p\big)
-\bar B_\mu\!\big(m_e,\,v\!\cdot\!k+v\!\cdot\!p\big)\Big],
\label{eq:r22-Bfunc}
\end{split}
\eeq
with the standard integrals (Appendix A of ref.~\cite{Eeg:2001un})
\begin{align}
i\,\mu^{2\epsilon}\!\int\!\frac{d^{4-2\epsilon}l}{(2\pi)^{4-2\epsilon}}
\frac{1}{l^2-m_e^2}
&= \frac{1}{16\pi^2}\,A_1(m_e),\\
i\,\mu^{2\epsilon}\!\int\!\frac{d^{4-2\epsilon}l}{(2\pi)^{4-2\epsilon}}
\frac{1}{(v\!\cdot\!l-\Delta)(l^2-m_e^2)}
&= \frac{1}{16\pi^2}\,\frac{1}{\Delta}\,\bar B_0(m_e,\Delta),\\
i\,\mu^{2\epsilon}\!\int\!\frac{d^{4-2\epsilon}l}{(2\pi)^{4-2\epsilon}}
\frac{l^\mu}{(v\!\cdot\!l-\Delta)(l^2-m_e^2)}
&= \frac{v^\mu}{16\pi^2}\,\big[\bar B_0(m_e,\Delta)+A_1(m_e)\big],
\end{align}
where
\begin{align}
A_1(m_e)&=m_e^2\log\!\frac{m_e^2}{\mu^2}
-m_e^2\Big(\frac{1}{\epsilon}-\gamma_E+\ln 4\pi+1\Big),\\
\bar B_0(m_e,\Delta)
&=-2\Delta^2\log\!\frac{m_e^2}{\mu^2}
-4\Delta^2 F\!\left(\frac{m_e}{\Delta}\right)
+2\Delta^2\Big(\frac{1}{\epsilon}-\gamma_E+\ln 4\pi+2\Big).
\end{align}
Collecting all the terms, gives
\begin{equation}
\mathcal M_{\langle r^2\rangle^2}
=\frac{i\,\alpha^2\,\langle r^2\rangle^2}{18}\,
\Big[(v\!\cdot\!k)^2-(v\!\cdot\!k')^2+m_e\,(v\!\cdot\!k-v\!\cdot\!k')\Big]\,
\Big[\frac{1}{\epsilon}-\gamma_E+2+\log\!\frac{4\pi\mu^2}{m_e^2}\Big],
\end{equation}
which vanishes in the static on-shell limit, $v \cdot k = v \cdot k' = m_e$.
Thus the $\langle r^2\rangle^2$ box and crossed-box diagrams do not generate a new long-range potential; any analytic remainder renormalizes local contact terms, consistent with the pNRQED matching logic \cite{Pineda:1998kn,Hill:2011wy,Hill:2012rh}.

\section{UV and IR divergencies} 
\label{Appendix:UV_IR_divergencies}  
The traditional approach to nuclear structure spectral corrections involves introducing a nuclear electric potential $\delta V(\bm r)$ parameterizing the deviation from a Coulomb potential \cite{Friar:1978wv}. In momentum space $\delta V(\bm r)$ is related to the electric form factor, the Fourier transform of the charge density in the Breit frame \cite{Halzen:1984mc}. The potential $\delta V(\bm r)$ is used in QM perturbation theory to express nuclear structure corrections to spectra in terms of moments of the charge density. In an EFT approach we express these corrections directly in terms of moments. In momentum space, this amounts to expanding the electric form factor in powers of $\bm q^2$. It is not surprising that this leads to UV divergencies since higher orders in the expansion introduce more and more powers of $\bm q^2$. In the main text we encountered this, e.g., in the contribution to the $s$-levels from the $\langle r^4 \rangle$ operator. To clarify this point, we present below a method of regions based analysis, see, e.g., section~2.1 of  \cite{Becher:2014oda}.

Define (sign convention follows \cite{Peskin:1995ev})
\begin{equation}
\delta V(\bm{r})=\int\dfrac{d^3\bm{q}}{(2\pi)^3}e^{i\bm{q}\cdot \bm{r}}\,\delta\tilde{V}(\bm{q})\equiv\int\dfrac{d^3\bm{q}}{(2\pi)^3}e^{i\bm{q}\cdot \bm{r}}\,\frac{\left(-4\pi\alpha Z\right)}{\bm{q}^{2}} \left[F(-\bm{q}^{2})-1\right],
\end{equation}
where $F(-\bm{q}^{2})$ is the electric form factor. We are working in the infinite nuclear mass limit where $q^2\approx -\bm{q}^{2}$. 
We consider the energy level shift induced by $\delta V$, in \emph{momentum} space: 
\begin{equation}\label{DEGeneral}
\Delta E=\int d^3 \bm{r}\,\, |\psi(\bm{r})|^2\,\delta V(\bm{r})=\int\dfrac{d^3\bm{q}}{(2\pi)^3}\,\frac{\left(-4\pi\alpha Z\right)}{\bm{q}^{2}} \left[F(-\bm{q}^{2})-1\right]\int d^3 \bm{r}\,e^{i\bm{q}\cdot \bm{r}}|\psi(\bm{r})|^2 \,.
\end{equation} 
To shorten the notation, we redefine for the remainder of this appendix, $F(-\bm{q}^{2})\to F(\bm{q}^{2})$. For simplicity, let us also focus only the energy shift of the hydrogen $1s$ level, where a similar analysis also applies to other $s$ levels.
The expression from the full theory is ($Q_e=-1$)
\begin{eqnarray}
\Delta E_{1s}^\textnormal{Full} &=& 4\pi\alpha ZQ _e (2m_eZ\alpha)^4 \int\dfrac{d^3\bm{q}}{(2\pi)^3} \frac{\left[F(\bm{q}^2)-1\right]}{\bm{q}^2(\bm{q}^2+(2m_eZ\alpha)^2)^2} \,.
\end{eqnarray} 
For the high-momentum region $|{\bm q}|^2\gg (m_eZ\alpha)^2$ we can approximate in the integrand $\bm{q}^2 + (2m_eZ\alpha)^2 \approx \bm{q}^2$:
 \begin{eqnarray}
\Delta E^\textnormal{High}_{1s} &=& 4\pi\alpha Z Q_e (2m_eZ\alpha)^4 \int_{|{\bm q}|>\Lambda} \dfrac{d^3\bm{q}}{(2\pi)^3} \frac{\left[F(\bm{q}^2)-1\right]}{\bm{q}^6 } \,.
\end{eqnarray} 
Without information on $F$ we cannot calculate the above integrals.

In the low-momentum region, $|{\bm q}|^2\ll (m_eZ\alpha)^2$, the integral can be expressed in terms of derivatives of $F(\bm{q}^2)$ with respect to $\bm{q}^2$ at $\bm{q}^2=0$. 
\beq
\begin{split}
\Delta E_{1s}^{\textnormal{Low, }F^\prime} &= 4\pi\alpha Z Q_e (2m_eZ\alpha)^4 \int_{|{\bm q}|\leq\Lambda} \dfrac{d^3\bm{q}}{(2\pi)^3} F^\prime(0) \frac{1}{(\bm{q}^2+(2m_eZ\alpha)^2)^2} \\
&= \frac{1}{2} (2m_eZ\alpha)^3\alpha Z Q_eF^\prime(0)-\frac{2(2m_eZ\alpha)^4\alpha Z Q_eF^\prime(0)}{\pi\Lambda}+\cdots \,,
\end{split} 
\eeq
and
\beq
\begin{split}
\Delta E_{1s}^{\textnormal{Low, }F^{\prime\prime}} &= 4\pi\alpha Z Q_e (2m_eZ\alpha)^4 \int_{|{\bm q}|\leq\Lambda} \dfrac{d^3\bm{q}}{(2\pi)^3} \frac{F^{\prime\prime}(0)}{2} \frac{\bm{q}^2}{(\bm{q}^2+(2m_eZ\alpha)^2)^2} \\
&=\frac{(2m_eZ\alpha)^4\alpha Z Q_eF^{\prime\prime}(0)\Lambda}{\pi}-\frac{3}{4} (2m_eZ\alpha)^5\alpha Z Q_eF^{\prime\prime}[0]+\cdots.
\end{split}
\eeq

Let's illustrate the method of regions analysis using the popular dipole model
\begin{equation}
F(\bm{q}^2)=\frac{\mphi^4}{(\bm{q}^{2}+\mphi^2)^2} \,.
\end{equation}

Using the dipole model for $F$ in the above expressions yield
\beq
\begin{split}
\label{eq:DeltaE:full}
\Delta E_{1s}^\textnormal{Full}&=-\frac{(2m_eZ\alpha)^3\alpha Z Q_e (m_eZ\alpha+\mphi)}{(2m_eZ\alpha+\mphi)^3} 
\\
&=-\frac{(2m_eZ\alpha)^3\alpha Z Q_e}{\mphi^2}+\frac{5}{2}\frac{(2m_eZ\alpha)^4\alpha Z Q_e}{\mphi^3}-\frac{9}{2} \frac{(2m_eZ\alpha)^5\alpha Z Q_e}{\mphi^4}+\cdots \,,
\end{split} 
\eeq
and
\beq
\begin{split}
\label{eq:DeltaE:high}
\Delta E^\textnormal{High}_{1s}&=
-\frac{4(2m_eZ\alpha)^4\alpha Z Q_e}{\mphi^2\pi\Lambda}+\frac{5}{2}\frac{(2m_eZ\alpha)^4\alpha Z Q_e}{\mphi^3}-\frac{6(2m_eZ\alpha)^4\alpha Z Q_e\Lambda}{\mphi^4\pi }+\cdots\,.
\end{split} 
\eeq
The first term in $\Delta E^\textnormal{High}_{1s}$ is IR divergent in the $\Lambda\to 0$ limit, while the second term is the $1/\mphi^3$  term in the full result, eq. \eqref{eq:DeltaE:full}. These terms mirror eq.~(S14) in \cite{Berengut:2017zuo}. Finally, the third term in eq. \eqref{eq:DeltaE:high} is finite in the $\Lambda\to 0$ limit. 

For the low-momentum region the $F^\prime(0)$ term gives 
\begin{eqnarray}
\Delta E_{1s}^{\textnormal{Low, }F^\prime} &=& -\,\frac{(2m_eZ\alpha)^3\alpha Z Q_e}{\mphi^2}+\frac{4(2m_eZ\alpha)^4\alpha Z Q_e}{\mphi^2\pi\Lambda}+\cdots\,.
\end{eqnarray} 
The first term is the $1/\mphi^2$  term in the full result. The second term is finite in the $\Lambda\to \infty$ limit. When we add all the regions together, it will cancel the IR divergence of the high momentum region. The $F^{\prime\prime}(0)$ term gives
\begin{eqnarray}
\Delta E_{1s}^{\textnormal{Low, }F^{\prime\prime}} &=& \frac{6(2m_eZ\alpha)^4\alpha Z Q_e\Lambda}{\mphi^4\pi}-\frac{9}{2} \frac{(2m_eZ\alpha)^5\alpha Z Q_e}{\mphi^4}+\dots\,.
\end{eqnarray} 
The first term is UV divergent in the $\Lambda\to \infty$ limit. When we add all the regions together, it will cancel against the ${\cal O}(\Lambda)$ term from the high momentum region. The second term is the $1/\mphi^4$  term in the full result.

Adding up the three integrals we find that $\Delta E_{1s}^\textnormal{High} + \Delta E_{1s}^{\textnormal{Low, }F^\prime} + \Delta E_{1s}^{\textnormal{Low, }F^{\prime\prime}} = \Delta E_{1s}^\textnormal{Full}$. 
The analysis confirms the conclusion of \cite{Berengut:2017zuo} that the $1/\mphi^3$ originates from the high-energy region.  
The analysis also shows that the UV divergence we encounter from $\nabla^2\delta^3(\bm{r})$ will be removed by high-momentum effects and thus can be ignored. The remaining finite result gives the correct answer.

\section{Further details on $\langle r^2\rangle$ induced shifts in the wave functions}
\label{sec:app:r22}  
Here we provide further details relevant for the  results in sec. \ref{sec:other:r22}. From \cite{Friar:1978wv} the difference in electronic density at the origin for $i=(n,\ell)\to (n',\ell')$ transition is given by
\beq
\label{eq:rho_i:def}
\rho_{i}(0;\langle r^2 \rangle)=(\phi_{n',\ell'} + \Delta \phi_{n',\ell'})^2-(\phi_{n,\ell} + \Delta \phi_{n,\ell})^2,
\eeq
where $\phi_{n,\ell}=\phi_{n,\ell}(0)$ is the value of the wave function at the origin for point-like nucleus, and  $\Delta \phi_{n,\ell}$ is the correction due to the composite nucleus.  

For hydrogen-like systems, the shift in the wave function entering \eqref{eq:rho_i:def} is given by (setting $\ell=\ell'=0$ for simplicity) \cite{Friar:1978wv}
\beq	
(\phi_n + \Delta \phi_n)^2 = |\phi_n(0)|^2 + \Delta \phi_n^2 + 2\,\phi_n\,\Delta \phi_n \,,
\eeq
where keeping only order $(Z\,\alpha\,m_e)^5$ terms, 
\begin{align}
    \begin{split}
    \phi_n \Delta \phi_n =& (Z\,\alpha\,m_e)|\phi_n(0)|^2 \Big\{-\langle r \rangle + \frac{2\,Z\,\alpha\,m_e}{3} \langle r^2 \rangle \,B(n) + \frac{2\,Z\,\alpha\,m_e}{3} \langle r^2 \rangle \,(2\,\gamma_E-4/3) 
    \\
    &\qquad\qquad\qquad\qquad+ \frac{2\,Z\,\alpha\,m_e}{3} \langle r^2 \log(2\,Z\,\alpha\,m_e\,r) \rangle \Big\}\,,
    \end{split}\\
    \begin{split}
    \Delta \phi_n^2 =& |\phi_n(0)|^2 (Z\,\alpha\,m_e)^2 
    \langle r \rangle^2\,,
\end{split}   
\end{align}
and $B(n) = \psi (n) - 1/n - \ln n$. For $nS \rightarrow n'S$ transition,
\beq
\begin{split}
\label{eq:rho0}
    \rho_{nS\rightarrow n'S}(0;\langle r^2 \rangle) =& \frac{(Z\,\alpha\,m_e)^3}{\pi}\biggr\{\biggr[1-2\,Z\,\alpha\,m_e\langle r\rangle + (Z\,\alpha\,m_e)^2 \Big( \frac{4}{3}\langle r^2 \log(2\,Z\,\alpha\,m_e\,r) \rangle \\
    &+\langle r \rangle^2\Big) 
    + 
    \frac{4}{3}(Z\,\alpha\,m_e)^2 \langle r^2 \rangle
    (2\,\gamma_E-4/3)\biggr]\left(\frac{1}{n'^3}-\frac{1}{n^3}\right) 
    \\
    &+ \biggr[\frac{4}{3}(Z\,\alpha\,m_e)^2 \langle r^2 \rangle \biggr]\left(\frac{B(n')}{n'^3}-\frac{B(n)}{n^3}\right)\biggr\}.
    \end{split}
\eeq

Notice that it depends not just on $\langle r^2 \rangle$ but also on other moments of the general (unknown) charge distribution: $\langle r\rangle$ and $\langle r^2 \log(2Z\alpha m_e r)\rangle$.

The partial derivative with respect to \(\langle r^2 \rangle\), keeping all other expectation values fixed, and working to order \((Z\alpha m_e)^5\), is given by
\begin{eqnarray}\label{eq:part_rho0}
    \frac{\partial \rho_{nS\rightarrow n'S}(0;\langle r^2 \rangle)}{\partial \langle r^2 \rangle} = \frac{4(Z\alpha m_e)^5}{3\pi}\left\{\left(2\,\gamma_E-\frac43\right)\left(\frac{1}{n'^3}-\frac{1}{n^{3}}\right)+\left(\frac{B(n')}{n'^3}-\frac{B(n)}{n^{3}}\right)\right\}.
\end{eqnarray}
This gives from eq.~\eqref{eq:G2i:short-cut},
\begin{equation}\label{eq:g2_approx_apx}
    G^{(2)\text{app}}_{nS\rightarrow n'S} = \frac{4 (Z\alpha)(Z\alpha m_e)^5}{9}\left\{\left(2\,\gamma_E-\frac43\right)\left(\frac{1}{n'^3}-\frac{1}{n^{3}}\right)+\left(\frac{B(n')}{n'^3}-\frac{B(n)}{n^{3}}\right)\right\},
\end{equation}
which differs by an extra factor proportional to  \((2\,\gamma_E-4/3)\) from the exact expression for \(G^{(2)}_{nS\rightarrow n'S}\) in eq.~\eqref{eq:IS_contributions_hydrogen},
\begin{equation}\label{eq:g2_apx}
    G^{(2)}_{nS\rightarrow n'S} = \frac{4 (Z\alpha)(Z\alpha m_e)^5}{9}\left\{\left(\frac{B(n')}{n'^3}-\frac{B(n)}{n^{3}}\right)\right\}\,.
\end{equation}

The density for the point charge nucleus is obtained from small\(-r\) expansion of the wavefunction \(\phi_n(r)\),
\begin{equation}
    \rho_i^P(r^2=\langle r^2 \rangle) = \left(|\phi_{n^\prime,\ell^\prime}(r)|^2-|\phi_{n,\ell}(r)|^2\right)\Bigg|_{r^2=\langle r^2 \rangle},
\end{equation}
where \cite{Friar:1978wv},
\begin{equation}
    |\phi_{n,\ell}(r)|^2 = |\phi_{n,\ell}(0)|^2\left\{1- (2Z\alpha m_e r) + (Z\alpha m_e r)^2+\frac{2(Z\alpha m_e r)^2}{3}\left(1+\frac{1}{2n^2}\right) \right\}\,.
\end{equation}
Hence for $nS \rightarrow n'S$ transition,
\beq
\begin{split}\label{eq:rhop}
\rho^P_{nS \rightarrow n'S}\Big|_{r^2= \langle r^2 \rangle} &= \frac{(Z\,\alpha\,m_e)^3}{\pi} \Bigg\{ \left[1- 2\,Z\,\alpha\,m_e \sqrt{\langle r^2 \rangle} + \frac{5}{3}(Z\,\alpha\,m_e)^2 \langle r^2 \rangle\right]\times 
\\
&\times \left(\frac{1}{n'^3}-\frac{1}{n^3}\right)+ \frac{(Z\,\alpha\,m_e)^2\langle r^2 \rangle}{3}\left(\frac{1}{n'^5}-\frac{1}{n^5}\right) \Bigg\}\,,
\end{split}
\eeq
and its partial derivative with respect to \(r^2\) is
\begin{eqnarray}\label{eq:part_rhop}
    \frac{\partial \rho^P_{nS \rightarrow n'S}}{\partial r^2 }\Bigg|_{r^2= \langle r^2 \rangle} = \frac{4(Z\alpha m_e)^5}{3\pi}\left\{ \frac{5}{4}\left(\frac{1}{n'^3}-\frac{1}{n^{3}}\right)+\frac{1}{4}\left(\frac{1}{n'^5}-\frac{1}{n^{5}}\right)\right\}\,.
\end{eqnarray}

For $n = 2$ and $n' = 3$ with $Z = 1$, $\alpha=1/137$, $r_p = 0.842$ fm, $m_e = 0.511$ MeV, we find from eq.~\eqref{eq:g2_approx_apx}, \(G^{(2)\text{app}}_{nS\rightarrow n'S}/h=-3.5\times10^{-5} \,\rm Hz\,\, fm^{-4}\) and from eq.~\eqref{eq:g2_apx} \(G^{(2)}_{nS\rightarrow n'S}/h=-2.9\times10^{-5}\,\rm Hz\,\,fm^{-4}\). Their ratio is $1.2$, since $G^{(2)\text{app}}_i$ includes also terms that are absorbed into $F_i$ in the King NL analysis. Note, however, that the numerical value for $G^{(2)\text{app}}$ depends crucially on how the other moments are treated. Here, we treated  $\langle r\rangle$ and $\langle r^2 \log (2 Z \alpha m_e r)\rangle$ as independent, while if one models the nucleus as a particular classical charge density, they would be correlated with $\langle r^2\rangle$. Evaluating numerically the derivative in eq.~\eqref{eq:part_rho0}, by varying the size of such classical charge density can thus pick up also parametrically larger contributions proportional to $\langle r\rangle$. 

Using the same values, $n=2,n'=3, Z=1$, the ratio of eq.~\eqref{eq:rho0} to eq.~\eqref{eq:rhop} gives \(C^P_{nS \rightarrow n'S} = 1\), while the ratio of eq.~\eqref{eq:part_rho0} to eq.~\eqref{eq:part_rhop} gives
\(C_{nS \rightarrow n'S} = -0.8\). The two are similar in magnitude, but have opposite signs. These results highlight the need for further study of how $G^{(2)}_i$ can be most efficiently evaluated.

\clearpage

\bibliographystyle{JHEP}
\bibliography{king_nl}

@article{Orce:2023qtm,
    author = "Orce, Jos{\'e} Nicol{\'a}s and Ngwetsheni, Cebo and Brown, B. Alex",
    title = "{Global trends of the electric dipole polarizability from shell-model calculations}",
    eprint = "2309.08810",
    archivePrefix = "arXiv",
    primaryClass = "nucl-th",
    doi = "10.1103/PhysRevC.108.044309",
    journal = "Phys. Rev. C",
    volume = "108",
    number = "4",
    pages = "044309",
    year = "2023"
}

@article{Lepage:1992tx,
    author = "Lepage, G. Peter and Magnea, Lorenzo and Nakhleh, Charles and Magnea, Ulrika and Hornbostel, Kent",
    title = "{Improved nonrelativistic QCD for heavy quark physics}",
    eprint = "hep-lat/9205007",
    archivePrefix = "arXiv",
    reportNumber = "CLNS-92-1136, OHSTPY-HEP-T-92-001",
    doi = "10.1103/PhysRevD.46.4052",
    journal = "Phys. Rev. D",
    volume = "46",
    pages = "4052--4067",
    year = "1992"
}

@article{Luke:1999kz,
    author = "Luke, Michael E. and Manohar, Aneesh V. and Rothstein, Ira Z.",
    title = "{Renormalization group scaling in nonrelativistic QCD}",
    eprint = "hep-ph/9910209",
    archivePrefix = "arXiv",
    reportNumber = "UCSD-PTH-99-11, UTPT-99-15, CMU-99-11",
    doi = "10.1103/PhysRevD.61.074025",
    journal = "Phys. Rev. D",
    volume = "61",
    pages = "074025",
    year = "2000"
}

@article{Carlson:2011zd,
    author = "Carlson, Carl E. and Vanderhaeghen, Marc",
    title = "{Higher order proton structure corrections to the Lamb shift in muonic hydrogen}",
    eprint = "1101.5965",
    archivePrefix = "arXiv",
    primaryClass = "hep-ph",
    reportNumber = "MKPH-T-11-03",
    doi = "10.1103/PhysRevA.84.020102",
    journal = "Phys. Rev. A",
    volume = "84",
    pages = "020102",
    year = "2011"
}

@article{Kobach:2017xkw,
    author = "Kobach, Andrew and Pal, Sridip",
    title = "{Hilbert Series and Operator Basis for NRQED and NRQCD/HQET}",
    eprint = "1704.00008",
    archivePrefix = "arXiv",
    primaryClass = "hep-ph",
    doi = "10.1016/j.physletb.2017.06.026",
    journal = "Phys. Lett. B",
    volume = "772",
    pages = "225--231",
    year = "2017"
}

@article{Borah:2020gte,
    author = "Borah, Kaushik and Hill, Richard J. and Lee, Gabriel and Tomalak, Oleksandr",
    title = "{Parametrization and applications of the low-$Q^2$ nucleon vector form factors}",
    eprint = "2003.13640",
    archivePrefix = "arXiv",
    primaryClass = "hep-ph",
    reportNumber = "FERMILAB-PUB-20-124-T",
    doi = "10.1103/PhysRevD.102.074012",
    journal = "Phys. Rev. D",
    volume = "102",
    number = "7",
    pages = "074012",
    year = "2020"
}

@article{Antognini:2022xoo,
   
author = "Antognini, Aldo and Hagelstein, Franziska and Pascalutsa, Vladimir",
    title = "{The proton structure in and out of muonic hydrogen}",
    eprint = "2205.10076",
    archivePrefix = "arXiv",
    primaryClass = "nucl-th",
    reportNumber = "MITP/22-039, PSI-PR-22-29",
    doi = "10.1146/annurev-nucl-101920-024709",
    journal = "Ann. Rev. Nucl. Part. Sci.",
    volume = "72",
    pages = "389",
    year = "2022"
}

@article{Hoang:2005dk,
    author = "Hoang, Andre H. and Ruiz-Femenia, Pedro",
    title = "{Renormalization group analysis in NRQCD for colored scalars}",
    eprint = "hep-ph/0511102",
    archivePrefix = "arXiv",
    reportNumber = "MPP-2005-130",
    doi = "10.1103/PhysRevD.73.014015",
    journal = "Phys. Rev. D",
    volume = "73",
    pages = "014015",
    year = "2006"
}

@article{Borisoglebsky:1979nb,
    author = "Borisoglebsky, L. A. and Trofimenko, E. E",
    title = "{The nuclear size correction to the energy levels of the light hydrogen-like and muonic atoms}",
    doi = "10.1016/0370-2693(79)90516-1",
    journal = "Phys. Lett. B",
    volume = "81",
    pages = "175--177",
    year = "1979"
}

@article{Berengut:2017zuo,
    author = "Berengut, Julian C. and others",
    title = "{Probing New Long-Range Interactions by Isotope Shift Spectroscopy}",
    eprint = "1704.05068",
    archivePrefix = "arXiv",
    primaryClass = "hep-ph",
    reportNumber = "DESY-17-055, FERMILAB-PUB-17-077-T, LAPTh-009-17, MIT-CTP-4898",
    doi = "10.1103/PhysRevLett.120.091801",
    journal = "Phys. Rev. Lett.",
    volume = "120",
    pages = "091801",
    year = "2018"
}

@article{Delaunay:2016brc,
    author = "Delaunay, C{\'e}dric and Ozeri, Roee and Perez, Gilad and Soreq, Yotam",
    title = "{Probing Atomic Higgs-like Forces at the Precision Frontier}",
    eprint = "1601.05087",
    archivePrefix = "arXiv",
    primaryClass = "hep-ph",
    doi = "10.1103/PhysRevD.96.093001",
    journal = "Phys. Rev. D",
    volume = "96",
    number = "9",
    pages = "093001",
    year = "2017"
}

@book{Halzen:1984mc,
    author = "Halzen, F. and Martin, Alan D.",
    title = "{QUARKS AND LEPTONS: AN INTRODUCTORY COURSE IN MODERN PARTICLE PHYSICS}",
    isbn = "978-0-471-88741-6",
    year = "1984"
}

@article{Luke:1992cs,
    author = "Luke, Michael E. and Manohar, Aneesh V.",
    title = "{Reparametrization invariance constraints on heavy particle effective field theories}",
    eprint = "hep-ph/9205228",
    archivePrefix = "arXiv",
    reportNumber = "UCSD-PTH-92-15",
    doi = "10.1016/0370-2693(92)91786-9",
    journal = "Phys. Lett. B",
    volume = "286",
    pages = "348--354",
    year = "1992"
}

@book{Becher:2014oda,
    author = "Becher, Thomas and Broggio, Alessandro and Ferroglia, Andrea",
    title = "{Introduction to Soft-Collinear Effective Theory}",
    eprint = "1410.1892",
    archivePrefix = "arXiv",
    primaryClass = "hep-ph",
    reportNumber = "PSI-PR-14-12",
    doi = "10.1007/978-3-319-14848-9",
    publisher = "Springer",
    volume = "896",
    year = "2015"
}

@misc{toappear,
  author    = {Beno\^{i}t Assi and Sam Carey and Sebastian J\"ager and Gabriel Lee and Gil Paz and Gilad Perez and Jure Zupan},
  title     = {},
  year      = {To appear},
  note      = {},
}

@article{Stryker:2015ika,
    author = "Stryker, Jesse R. and Miller, Gerald A.",
    title = "{Proton charge extensions}",
    eprint = "1508.06680",
    archivePrefix = "arXiv",
    primaryClass = "hep-ph",
    reportNumber = "NT@UW-15-11",
    doi = "10.1103/PhysRevA.93.012509",
    journal = "Phys. Rev. A",
    volume = "93",
    number = "1",
    pages = "012509",
    year = "2016"
}

@inproceedings{Lepage:1997cs,
    author = "Lepage, G. P.",
    title = "{How to renormalize the Schrodinger equation}",
    booktitle = "{8th Jorge Andre Swieca Summer School on Nuclear Physics}",
    eprint = "nucl-th/9706029",
    archivePrefix = "arXiv",
    pages = "135--180",
    month = "2",
    year = "1997"
}

@article{King:1963,
  title={Comments on the article ``peculiarities of the isotope shift in the samarium spectrum''},
  author={King, WH},
  journal={Journal of the Optical Society of America},
  volume={53},
  number={5},
  pages={638--639},
  year={1963},
  publisher={Optical Society of America}
}

@article{Flambaum:2017onb,
    author = "Flambaum, V. V. and Geddes, A. J. and Viatkina, A. V.",
    title = "{Isotope shift, nonlinearity of King plots, and the search for new particles}",
    eprint = "1709.00600",
    archivePrefix = "arXiv",
    primaryClass = "physics.atom-ph",
    doi = "10.1103/PhysRevA.97.032510",
    journal = "Phys. Rev. A",
    volume = "97",
    number = "3",
    pages = "032510",
    year = "2018"
}

@article{Frugiuele:2016rii,
    author = "Frugiuele, Claudia and Fuchs, Elina and Perez, Gilad and Schlaffer, Matthias",
    title = "{Constraining New Physics Models with Isotope Shift Spectroscopy}",
    eprint = "1602.04822",
    archivePrefix = "arXiv",
    primaryClass = "hep-ph",
    doi = "10.1103/PhysRevD.96.015011",
    journal = "Phys. Rev. D",
    volume = "96",
    number = "1",
    pages = "015011",
    year = "2017"
}

@article{Counts:2020aws,
    author = "Counts, Ian and Hur, Joonseok and Aude Craik, Diana P. L. and Jeon, Honggi and Leung, Calvin and Berengut, Julian C. and Geddes, Amy and Kawasaki, Akio and Jhe, Wonho and Vuleti\'c, Vladan",
    title = "{Evidence for Nonlinear Isotope Shift in Yb$^+$ Search for New Boson}",
    eprint = "2004.11383",
    archivePrefix = "arXiv",
    primaryClass = "physics.atom-ph",
    doi = "10.1103/PhysRevLett.125.123002",
    journal = "Phys. Rev. Lett.",
    volume = "125",
    number = "12",
    pages = "123002",
    year = "2020"
}

@article{Hur:2022gof,
    author = "Hur, Joonseok and others",
    title = "{Evidence of Two-Source King Plot Nonlinearity in Spectroscopic Search for New Boson}",
    eprint = "2201.03578",
    archivePrefix = "arXiv",
    primaryClass = "physics.atom-ph",
    doi = "10.1103/PhysRevLett.128.163201",
    journal = "Phys. Rev. Lett.",
    volume = "128",
    number = "16",
    pages = "163201",
    year = "2022"
}

@article{Ono:2021ogd,
    author = "Ono, Koki and Saito, Yugo and Ishiyama, Taiki and Higomoto, Toshiya and Takano, Tetsushi and Takasu, Yosuke and Yamamoto, Yasuhiro and Tanaka, Minoru and Takahashi, Yoshiro",
    title = "{Observation of Nonlinearity of Generalized King Plot in the Search for New Boson}",
    eprint = "2110.13544",
    archivePrefix = "arXiv",
    primaryClass = "physics.atom-ph",
    reportNumber = "OU-HET-1109",
    doi = "10.1103/PhysRevX.12.021033",
    journal = "Phys. Rev. X",
    volume = "12",
    number = "2",
    pages = "021033",
    year = "2022"
}

@article{Door:2024qqz,
    author = "Door, Menno and others",
    title = "{Probing New Bosons and Nuclear Structure with Ytterbium Isotope Shifts}",
    eprint = "2403.07792",
    archivePrefix = "arXiv",
    primaryClass = "physics.atom-ph",
    doi = "10.1103/PhysRevLett.134.063002",
    journal = "Phys. Rev. Lett.",
    volume = "134",
    number = "6",
    pages = "063002",
    year = "2025"
}

@article{Rehbehn:2021zlr,
    author = "Rehbehn, Nils-Holger and others",
    title = "{Sensitivity to New Physics of Isotope Shift Studies using the Coronal Lines of Highly Charged Calcium Ions}",
    eprint = "2102.02309",
    archivePrefix = "arXiv",
    primaryClass = "physics.atom-ph",
    doi = "10.1103/PhysRevA.103.L040801",
    journal = "Phys. Rev. A",
    volume = "103",
    number = "4",
    pages = "L040801",
    year = "2021"
}

@article{Wilzewski:2024wap,
    author = "Wilzewski, Alexander and others",
    title = "{Nonlinear Calcium King Plot Constrains New Bosons and Nuclear Properties}",
    eprint = "2412.10277",
    archivePrefix = "arXiv",
    primaryClass = "physics.atom-ph",
    doi = "10.1103/PhysRevLett.134.233002",
    journal = "Phys. Rev. Lett.",
    volume = "134",
    number = "23",
    pages = "233002",
    year = "2025"
}

@article{Pineda:1998kn,
    author = "Pineda, A. and Soto, J.",
    title = "{Potential NRQED: The Positronium case}",
    eprint = "hep-ph/9805424",
    archivePrefix = "arXiv",
    reportNumber = "UB-ECM-PF-98-11, KFA-IKP-TH-98-9",
    doi = "10.1103/PhysRevD.59.016005",
    journal = "Phys. Rev. D",
    volume = "59",
    pages = "016005",
    year = "1999"
}

@article{Ong:2010gf,
    author = "Ong, A. and Berengut, J. C. and Flambaum, V. V.",
    title = "{The Effect of spin-orbit nuclear charge density corrections due to the anomalous magnetic moment on halonuclei}",
    eprint = "1006.5508",
    archivePrefix = "arXiv",
    primaryClass = "nucl-th",
    doi = "10.1103/PhysRevC.82.014320",
    journal = "Phys. Rev. C",
    volume = "82",
    pages = "014320",
    year = "2010"
}

@article{Caputo:2024doz,
    author = "Caputo, Andrea and Gazit, Doron and Hammer, Hans-Werner and Kopp, Joachim and Paz, Gil and Perez, Gilad and Springmann, Konstantin",
    title = "{Sensitivity of nuclear clocks to new physics}",
    eprint = "2407.17526",
    archivePrefix = "arXiv",
    primaryClass = "hep-ph",
    reportNumber = "CERN-TH-2024-117, MITP-24-063, TUM-HEP-1515/24, WSU-HEP-2403",
    doi = "10.1103/l29n-gt5j",
    journal = "Phys. Rev. C",
    volume = "112",
    number = "3",
    pages = "L031302",
    year = "2025"
}

@article{ParticleDataGroup:2024cfk,
    author = "Navas, S. and others",
    collaboration = "Particle Data Group",
    title = "{Review of particle physics}",
    doi = "10.1103/PhysRevD.110.030001",
    journal = "Phys. Rev. D",
    volume = "110",
    number = "3",
    pages = "030001",
    year = "2024"
}

@article{Angeli:2013epw,
    author = "Angeli, I. and Marinova, K. P.",
    title = "{Table of experimental nuclear ground state charge radii: An update}",
    doi = "10.1016/j.adt.2011.12.006",
    journal = "Atom. Data Nucl. Data Tabl.",
    volume = "99",
    number = "1",
    pages = "69--95",
    year = "2013"
}

@article{Dye:2016uep,
    author = "Dye, Steven P. and Gonderinger, Matthew and Paz, Gil",
    title = "{Elements of QED-NRQED effective field theory: NLO scattering at leading power}",
    eprint = "1602.07770",
    archivePrefix = "arXiv",
    primaryClass = "hep-ph",
    reportNumber = "WSU-HEP-1601",
    doi = "10.1103/PhysRevD.94.013006",
    journal = "Phys. Rev. D",
    volume = "94",
    number = "1",
    pages = "013006",
    year = "2016"
}

@article{Dye:2018rgg,
    author = "Dye, Steven P. and Gonderinger, Matthew and Paz, Gil",
    title = "{Elements of QED-NRQED Effective Field Theory: II. Matching of Contact Interactions}",
    eprint = "1812.05056",
    archivePrefix = "arXiv",
    primaryClass = "hep-ph",
    reportNumber = "FERMILAB-PUB-18-675-T, WSU-HEP-1808",
    doi = "10.1103/PhysRevD.100.054010",
    journal = "Phys. Rev. D",
    volume = "100",
    number = "5",
    pages = "054010",
    year = "2019"
}

@article{Eides:2000xc,
    author = "Eides, Michael I. and Grotch, Howard and Shelyuto, Valery A.",
    title = "{Theory of light hydrogen - like atoms}",
    eprint = "hep-ph/0002158",
    archivePrefix = "arXiv",
    reportNumber = "PSU-TH-226",
    doi = "10.1016/S0370-1573(00)00077-6",
    journal = "Phys. Rept.",
    volume = "342",
    pages = "63--261",
    year = "2001"
}

@article{Friar:1978wv,
    author = "Friar, James Lewis",
    title = "{Nuclear Finite Size Effects in Light Muonic Atoms}",
    reportNumber = "LA-UR-78-3093",
    doi = "10.1016/0003-4916(79)90300-2",
    journal = "Annals Phys.",
    volume = "122",
    pages = "151",
    year = "1979"
}

@article{Gunawardana:2017zix,
    author = "Gunawardana, Ayesh and Paz, Gil",
    title = "{On HQET and NRQCD Operators of Dimension 8 and Above}",
    eprint = "1702.08904",
    archivePrefix = "arXiv",
    primaryClass = "hep-ph",
    reportNumber = "WSU-HEP-1701",
    doi = "10.1007/JHEP07(2017)137",
    journal = "JHEP",
    volume = "07",
    pages = "137",
    year = "2017"
}

@article{Hill:2011wy,
    author = "Hill, Richard J. and Paz, Gil",
    title = "{Model independent analysis of proton structure for hydrogenic bound states}",
    eprint = "1103.4617",
    archivePrefix = "arXiv",
    primaryClass = "hep-ph",
    reportNumber = "EFI-PREPRINT-11-9",
    doi = "10.1103/PhysRevLett.107.160402",
    journal = "Phys. Rev. Lett.",
    volume = "107",
    pages = "160402",
    year = "2011"
}

@article{Hill:2011be,
    author = "Hill, Richard J. and Solon, Mikhail P.",
    title = "{Universal behavior in the scattering of heavy, weakly interacting dark matter on nuclear targets}",
    eprint = "1111.0016",
    archivePrefix = "arXiv",
    primaryClass = "hep-ph",
    doi = "10.1016/j.physletb.2012.01.013",
    journal = "Phys. Lett. B",
    volume = "707",
    pages = "539--545",
    year = "2012"
}

@article{Hill:2012rh,
    author = "Hill, Richard J. and Lee, Gabriel and Paz, Gil and Solon, Mikhail P.",
    title = "{NRQED Lagrangian at order $1/M^4$}",
    eprint = "1212.4508",
    archivePrefix = "arXiv",
    primaryClass = "hep-ph",
    reportNumber = "EFI-PREPRINT-12-8",
    doi = "10.1103/PhysRevD.87.053017",
    journal = "Phys. Rev. D",
    volume = "87",
    pages = "053017",
    year = "2013"
}

@article{Heinonen:2012km,
    author = "Heinonen, Johannes and Hill, Richard J. and Solon, Mikhail P.",
    title = "{Lorentz invariance in heavy particle effective theories}",
    eprint = "1208.0601",
    archivePrefix = "arXiv",
    primaryClass = "hep-ph",
    reportNumber = "EFI-PREPRINT-12-16",
    doi = "10.1103/PhysRevD.86.094020",
    journal = "Phys. Rev. D",
    volume = "86",
    pages = "094020",
    year = "2012"
}

@article{Hur:2022,
    author = "Hur, J. and others",
    title = "{Evidence of Two-Source King Plot Nonlinearity in Spectroscopic Search for New Boson}",
    eprint = "2201.03578",
    archivePrefix = "arXiv",
    doi = "10.1103/PhysRevLett.128.163201",
    journal = "Phys. Rev. Lett",
    volume = "128",
    pages = "163201",
    year = "2022"
}

@article{Kinoshita:1995mt,
    author = "Kinoshita, T. and Nio, M.",
    title = "{Radiative corrections to the muonium hyperfine structure. 1. The alpha**2 (Z-alpha) correction}",
    eprint = "hep-ph/9512327",
    archivePrefix = "arXiv",
    reportNumber = "CLNS-95-1382",
    doi = "10.1103/PhysRevD.53.4909",
    journal = "Phys. Rev. D",
    volume = "53",
    pages = "4909--4929",
    year = "1996"
}

@article{Manohar:1997qy,
    author = "Manohar, Aneesh V.",
    title = "{The HQET / NRQCD Lagrangian to order alpha / m-3}",
    eprint = "hep-ph/9701294",
    archivePrefix = "arXiv",
    reportNumber = "UCSD-PTH-97-01",
    doi = "10.1103/PhysRevD.56.230",
    journal = "Phys. Rev. D",
    volume = "56",
    pages = "230--237",
    year = "1997"
}

@article{Szafron:2019tho,
    author = "Szafron, Robert and Korzinin, Evgeny Yu. and Shelyuto, Valery A. and Ivanov, Vladimir G. and Karshenboim, Savely G.",
    title = {{Virtual Delbr{\"u}ck scattering and the Lamb shift in light hydrogenlike atoms}},
    eprint = "1909.04116",
    archivePrefix = "arXiv",
    primaryClass = "physics.atom-ph",
    reportNumber = "TUM-HEP-1209/19",
    doi = "10.1103/PhysRevA.100.032507",
    journal = "Phys. Rev. A",
    volume = "100",
    number = "3",
    pages = "032507",
    year = "2019"
}

@article{Feinberg:1989ps,
    author = "Feinberg, G. and Sucher, J. and Au, C. K.",
    title = "{The Dispersion Theory of Dispersion Forces}",
    reportNumber = "CU-TP-437",
    doi = "10.1016/0370-1573(89)90111-7",
    journal = "Phys. Rept.",
    volume = "180",
    pages = "83",
    year = "1989"
}

@article{Ghosh:2024ctv,
    author = "Ghosh, Mitrajyoti and Grossman, Yuval and Sieng, Chinhsan and Yu, Bingrong",
    title = "{The neutrino force at all length scales}",
    eprint = "2410.19059",
    archivePrefix = "arXiv",
    primaryClass = "hep-ph",
    month = "10",
    year = "2024"
}

@article{Khriplovich:1997fi,
    author = "Khriplovich, I. B. and Sen'kov, R. A.",
    title = "{Proton polarizability contribution to hydrogen Lamb shift}",
    eprint = "nucl-th/9704043",
    archivePrefix = "arXiv",
    month = "4",
    year = "1997"
}

@article{Ericson:1972nhh,
    author = "Ericson, Torleif Erik Oskar and Huefner, J.",
    title = "{Theory of polarization shifts in exotic atoms}",
    doi = "10.1016/0550-3213(72)90111-3",
    journal = "Nucl. Phys. B",
    volume = "47",
    pages = "205--239",
    year = "1972"
}

@article{Migdal:1945,
    author = "Migdal, A. B.",
    title = "{Quadrupole and dipole $\gamma$-radiation of nuclei}",
    journal = "J. Exp. Theor. Phys.",
    volume = "15",
    pages = "81",
    year = "1945"
}

@article{Levinger:1957zz,
    author = "Levinger, J. S.",
    title = "{Migdal's and Khokhlov's Calculations of the Nuclear Photoeffect}",
    doi = "10.1103/PhysRev.107.554",
    journal = "Phys. Rev.",
    volume = "107",
    pages = "554--558",
    year = "1957"
}

@article{Liu:2025ows,
    author = "Liu, Hongkai and Ohayon, Ben and Shtaif, Omer and Soreq, Yotam",
    title = "{Probing new hadronic forces with heavy exotic atoms}",
    eprint = "2502.03537",
    archivePrefix = "arXiv",
    primaryClass = "hep-ph",
    month = "2",
    year = "2025"
}

@book{Peskin:1995ev,
    author = "Peskin, Michael E. and Schroeder, Daniel V.",
    title = "{An Introduction to quantum field theory}",
    doi = "10.1201/9780429503559",
    isbn = "978-0-201-50397-5, 978-0-429-50355-9, 978-0-429-49417-8",
    publisher = "Addison-Wesley",
    address = "Reading, USA",
    year = "1995"
}

@article{Eeg:2001un,
    author = "Eeg, J. O. and Fajfer, S. and Zupan, J.",
    title = "{Nonfactorizable contributions to the decay mode D0 ---{\ensuremath{>}} K0 anti-K0}",
    eprint = "hep-ph/0101215",
    archivePrefix = "arXiv",
    doi = "10.1103/PhysRevD.64.034010",
    journal = "Phys. Rev. D",
    volume = "64",
    pages = "034010",
    year = "2001"
}

@article{Mertig:1990an,
    author = "Mertig, R. and Bohm, M. and Denner, Ansgar",
    title = "{FEYN CALC: Computer algebraic calculation of Feynman amplitudes}",
    reportNumber = "PRINT-90-0639 (WURZBURG)",
    doi = "10.1016/0010-4655(91)90130-D",
    journal = "Comput. Phys. Commun.",
    volume = "64",
    pages = "345--359",
    year = "1991"
}

@article{Shtabovenko:2016sxi,
    author = "Shtabovenko, Vladyslav and Mertig, Rolf and Orellana, Frederik",
    title = "{New Developments in FeynCalc 9.0}",
    eprint = "1601.01167",
    archivePrefix = "arXiv",
    primaryClass = "hep-ph",
    reportNumber = "TUM-EFT-71-15",
    doi = "10.1016/j.cpc.2016.06.008",
    journal = "Comput. Phys. Commun.",
    volume = "207",
    pages = "432--444",
    year = "2016"
}

@article{Shtabovenko:2020gxv,
    author = "Shtabovenko, Vladyslav and Mertig, Rolf and Orellana, Frederik",
    title = "{FeynCalc 9.3: New features and improvements}",
    eprint = "2001.04407",
    archivePrefix = "arXiv",
    primaryClass = "hep-ph",
    reportNumber = "P3H-20-002, TTP19-020, TUM-EFT 130/19",
    doi = "10.1016/j.cpc.2020.107478",
    journal = "Comput. Phys. Commun.",
    volume = "256",
    pages = "107478",
    year = "2020"
}

@book{Manohar:2000dt,
    author = "Manohar, Aneesh V. and Wise, Mark B.",
    title = "{Heavy quark physics}",
    doi = "10.1017/9781009402125",
    isbn = "978-0-521-03757-0, 978-1-009-40212-5",
    volume = "10",
    year = "2000"
}

@article{Mack:1956xe,
    author = "Mack, J. E. and Arroe, H.",
    title = "{Isotope shift in atomic spectra}",
    doi = "10.1146/annurev.ns.06.120156.001001",
    journal = "Ann. Rev. Nucl. Part. Sci.",
    volume = "6",
    pages = "117--128",
    year = "1956"
}

@book{King1984,
  author    = {W. H. King},
  title     = {Isotope Shifts in Atomic Spectra},
  publisher = {Springer},
  year      = {1984},
  series    = {Physics of Atoms and Molecules},
  address   = {New York},
  isbn      = {978-0-306-41562-3}
}

@article{Fortson1990_NuclearStructurePNC,
  author  = {E. N. Fortson and Y. Pang and L. Wilets},
  title   = {Nuclear-Structure Effects in Atomic Parity Nonconservation},
  journal = {Phys. Rev. Lett.},
  year    = {1990},
  volume  = {65},
  pages   = {2857},
  doi     = {10.1103/PhysRevLett.65.2857}
}

@article{Solaro:2020dxz,
    author = "Solaro, Cyrille and Meyer, Steffen and Fisher, Karin and Berengut, Julian C. and Fuchs, Elina and Drewsen, Michael",
    title = "{Improved isotope-shift-based bounds on bosons beyond the Standard Model through measurements of the $^2$D$_{3/2} - ^2$D$_{5/2}$ interval in Ca$^+$}",
    eprint = "2005.00529",
    archivePrefix = "arXiv",
    primaryClass = "physics.atom-ph",
    reportNumber = "APS/123-QED, EFI-20-6, FERMILAB-PUB-20-142-T",
    doi = "10.1103/PhysRevLett.127.029901",
    journal = "Phys. Rev. Lett.",
    volume = "125",
    number = "12",
    pages = "123003",
    year = "2020",
    note = "[Erratum: Phys.Rev.Lett. 127, 029901 (2021)]"
}

@article{Berengut:2025nxp,
    author = "Berengut, Julian C. and Delaunay, C{\'e}dric",
    title = "{Precision isotope-shift spectroscopy for new physics searches and nuclear insights}",
    doi = "10.1038/s42254-024-00793-2",
    journal = "Nature Rev. Phys.",
    volume = "7",
    number = "2",
    pages = "119--125",
    year = "2025"
}

@article{Schumacher:2005an,
    author = "Schumacher, Martin",
    title = "{Polarizability of the nucleon and Compton scattering}",
    eprint = "hep-ph/0501167",
    archivePrefix = "arXiv",
    reportNumber = "HEP-PH-0501167",
    doi = "10.1016/j.ppnp.2005.01.033",
    journal = "Prog. Part. Nucl. Phys.",
    volume = "55",
    pages = "567--646",
    year = "2005"
}

\end{document}